\documentclass[sigconf]{acmart}

\usepackage{multirow}
\usepackage{array}
\usepackage{amsthm}
\newtheorem{definition}{Definition}[section]
\usepackage{subfig}
\usepackage{balance}

\AtBeginDocument{%
  \providecommand\BibTeX{{%
    \normalfont B\kern-0.5em{\scshape i\kern-0.25em b}\kern-0.8em\TeX}}}

\setcopyright{acmcopyright}

\copyrightyear{2024}
\acmYear{2024}
\setcopyright{acmlicensed}
\acmConference[WSDM '24]{Proceedings of the 17th ACM International Conference on Web Search and Data Mining}{March 4--8, 2024}{Merida, Mexico}
\acmBooktitle{Proceedings of the 17th ACM International Conference on Web Search and Data Mining (WSDM '24), March 4--8, 2024, Merida, Mexico}
\acmPrice{15.00}
\acmISBN{979-8-4007-0371-3/24/03}
\acmDOI{10.1145/3616855.3635787}

\begin{document}

\title{Collaboration and Transition: Distilling Item Transitions into Multi-Query Self-Attention for Sequential Recommendation}



\author{Tianyu Zhu}
\authornote{This work was done at Université de Montréal.}
\affiliation{%
  \institution{Beihang University}
  \city{Beijing}
  \country{China}
}
\email{ztybuaa@126.com}

\author{Yansong Shi}
\affiliation{%
  \institution{Tsinghua University}
  \city{Beijing}
  \country{China}
}
\email{shiys.18@sem.tsinghua.edu.cn}

\author{Yuan Zhang}
\affiliation{%
  \institution{Kuaishou Technology}
  \city{Beijing}
  \country{China}
}
\email{yuanz.pku@gmail.com}

\author{Yihong Wu}
\affiliation{%
  \institution{Université de Montréal}
  \city{Montréal}
  \country{Canada}
}
\email{yihong.wu@umontreal.ca}

\author{Fengran Mo}
\affiliation{%
  \institution{Université de Montréal}
  \city{Montréal}
  \country{Canada}
}
\email{fengran.mo@umontreal.ca}

\author{Jian-Yun Nie}
\affiliation{%
  \institution{Université de Montréal}
  \city{Montréal}
  \country{Canada}
}
\email{nie@iro.umontreal.ca}

\renewcommand{\shortauthors}{Tianyu Zhu et al.}
\begin{abstract}
Modern recommender systems employ various sequential modules such as self-attention to learn dynamic user interests. However, these methods are less effective in capturing collaborative and transitional signals within user interaction sequences. First, the self-attention architecture uses the embedding of a single item as the attention query, making it challenging to capture collaborative signals. Second, these methods typically follow an auto-regressive framework, which is unable to learn global item transition patterns. To overcome these limitations, we propose a new method called Multi-Query Self-Attention with Transition-Aware Embedding Distillation (MQSA-TED). First, we propose an $L$-query self-attention module that employs flexible window sizes for attention queries to capture collaborative signals. In addition, we introduce a multi-query self-attention method that balances the bias-variance trade-off in modeling user preferences by combining long and short-query self-attentions. Second, we develop a transition-aware embedding distillation module that distills global item-to-item transition patterns into item embeddings, which enables the model to memorize and leverage transitional signals and serves as a calibrator for collaborative signals. Experimental results on four real-world datasets demonstrate the effectiveness of the proposed modules.
\end{abstract}





\maketitle
\section{Introduction}

\begin{figure}[!t]
    \centering
    \subfloat{\includegraphics[width=1.5in]{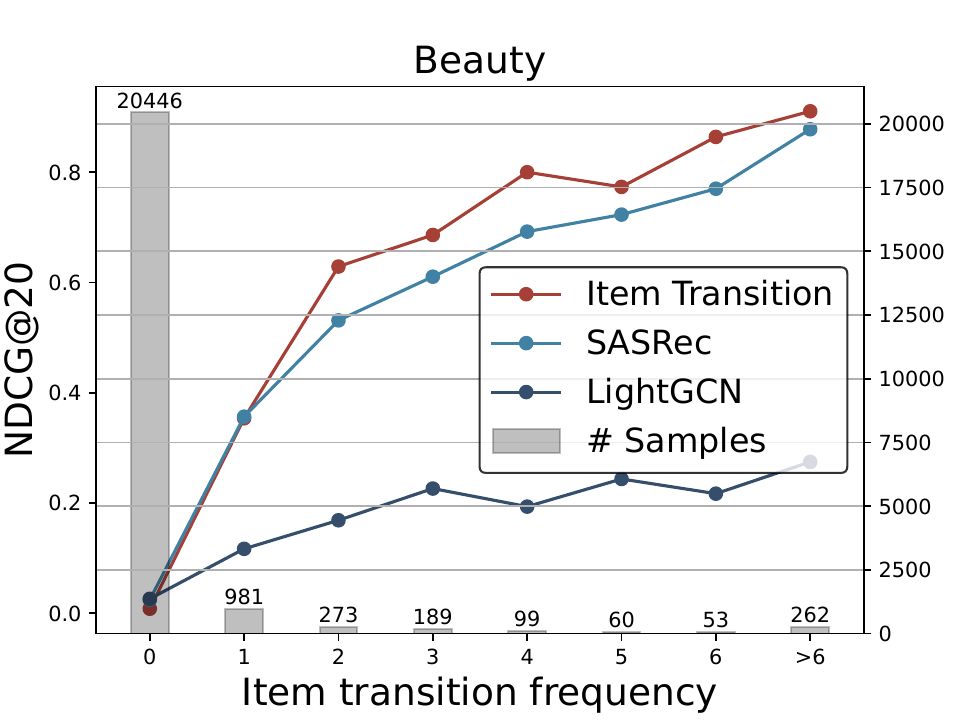}}
    \hfil
    \subfloat{\includegraphics[width=1.5in]{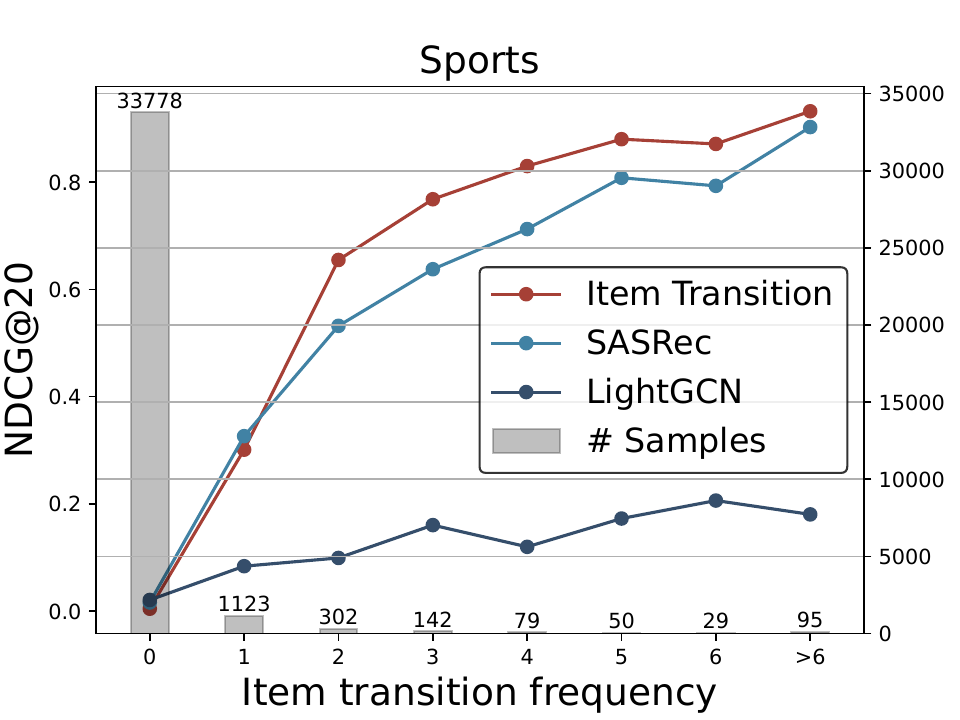}}
    \hfil
    \subfloat{\includegraphics[width=3.0in]{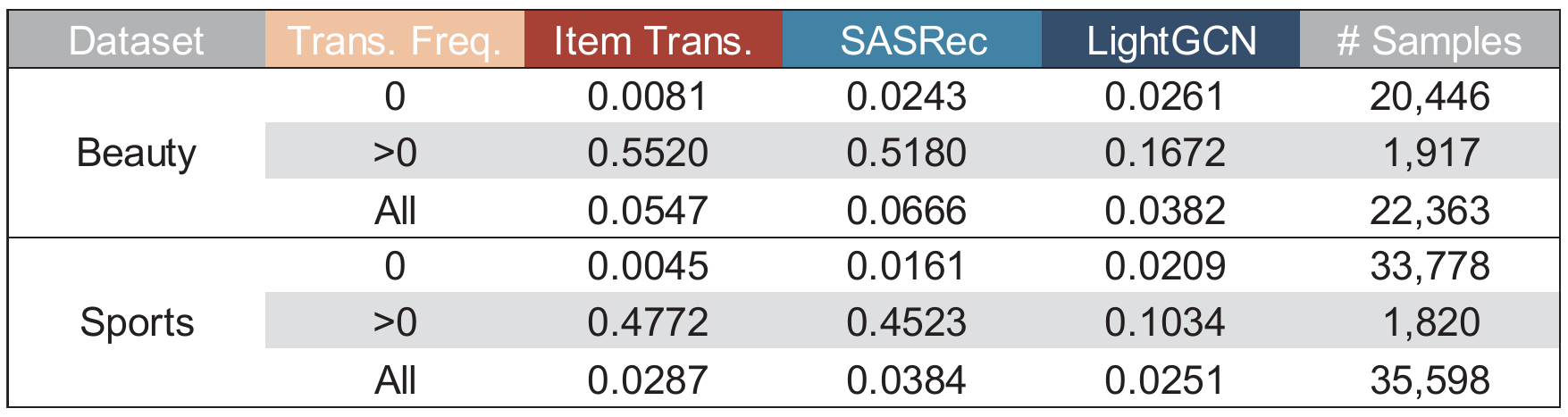}}
    \caption{Performance of three methods w.r.t. item transition frequency on two datasets. Item Transition performs better on test samples with frequent transitions, while LightGCN performs better on test samples lacking transition instances. SASRec achieves the best performance on average.}
    \label{fig:group_intro}
\end{figure}

In recent years, there has been an increasing focus on modeling dynamic user preferences in modern recommender systems \cite{zhou2018deep,gao2022kuairand}, which is achieved by incorporating various sequential modules such as RNN \cite{hidasi2015session}, CNN \cite{tang2018personalized}, and Transformer \cite{kang2018self,sun2019bert4rec}. These sequential recommenders aim to integrate contextual factors derived from recent user interactions into personalized user interests. Contextual factors reveal typical item-to-item transition patterns. The main challenge in sequential recommendation lies in effectively learning both personalized user interests and general item transition patterns while maintaining an appropriate balance between the two factors. For instance, a user interested in sportswear may also seek a shirt after purchasing a suit. If we only rely on collaborative signals to generate recommendations, we may overlook the user's temporary need for items to complement their suit. On the other hand, if we only consider transitional signals to make recommendations, we may neglect the user's primary interest in sportswear. Therefore, it is crucial to leverage both signals and find a balance between them. We define the collaborative and transitional signals in the context of sequential recommendation tasks as follows:

\begin{definition}[Collaborative Signals]
In the context of sequential recommendation, collaborative signals refer to the similarities between sequences of users' interacted items.
\end{definition}

\begin{definition}[Transitional Signals]
In the context of sequential recommendation, transitional signals refer to the transition frequency between pairs of users' interacted items.
\end{definition}

Specifically, collaborative signals can be used by following a sequence-to-item methodology, leveraging the collaborative behavior of users to identify patterns in their interactions and recommend relevant items. On the other hand, transitional signals exploit item-to-item relationships in user interaction sequences, enabling the identification of trigger items that will lead to related purchases.

Although recent sequential recommendation methods such as SASRec \cite{kang2018self} have demonstrated remarkable performance, they have inherent limitations in effectively capturing both signals within user interaction sequences. To highlight these limitations, we conducted experiments comparing the performance of SASRec with two baseline methods: Item Transition and LightGCN \cite{he2020lightgcn}. Item Transition is a memory-based, non-personalized method that makes recommendations based on the global transition frequency from the current item to candidate items, serving as a benchmark based on transitional signals (see Section 3.2 for details). LightGCN is a state-of-the-art non-sequential recommendation method that learns user and item embeddings through linear propagation on the user-item interaction graph, serving as a benchmark based on collaborative signals. We conducted experiments on two Amazon datasets, Beauty and Sports \cite{zhou2022filter}, and grouped the test samples based on the transition frequency observed in the training data. Results shown in Figure~\ref{fig:group_intro} reveal two limitations of SASRec in leveraging both signals:

First, SASRec has a lower ability to leverage collaborative signals than LightGCN. For test samples where the item transition frequency is zero, LightGCN consistently outperforms SASRec on both datasets. This observation shows the limited ability of SASRec to generalize to test samples lacking observed item transitions. Notably, SASRec uses the embedding of the most recent item as the query in its self-attention module, which can be regarded as an attention-enhanced first-order Markov chain model that is inherently limited in leveraging collaborative signals.

Second, SASRec's ability to leverage transitional signals is lower than Item Transition. For test samples where the item transition frequency exceeds one, i.e., the transition occurs multiple times in the training data, Item Transition significantly outperforms SASRec on both datasets. This observation highlights the limited effectiveness of SASRec in leveraging transitional signals.

Inspired by these observations, we propose a new method called \textit{Multi-Query Self-Attention with Transition-Aware Embedding Distillation} (MQSA-TED) for sequential recommendation tasks, which consists of two main components to capture collaborative and transitional signals, respectively. First, we propose an $L$-query self-attention module that uses several items (instead of a single item) in windows of flexible sizes as attention queries to capture collaborative signals. By enlarging the window size $L$, the model can leverage similarities between longer-range sequences of users' interacted items to generate recommendations. However, using a large $L$ will result in a global bias as the recommendation will mainly focus on the user's long-term interests while ignoring the interest shift over time. To strike a balance between bias and variance in modeling users' dynamic interests, we introduce a multi-query self-attention method by combining long and short-query self-attentions. Second, we develop a transition-aware embedding distillation module that distills global item-to-item transition patterns into item embeddings, which serves as a calibration module that enables the model to effectively memorize and leverage transitional signals when making recommendations. Notably, our proposed method achieves inherent disentanglement of user collaboration modeling and item transition modeling by employing dual supervision: the original item embedding captures item-to-item transitional signals, while the item embedding created after self-attention modules captures sequence-to-item collaborative signals. Our contributions in this paper are summarized as follows:

\begin{itemize}
    \item We propose an $L$-query self-attention module that uses flexible window sizes for attention queries to capture collaborative signals. We also design a multi-query self-attention method that combines long and short-query self-attentions to balance the bias-variance trade-off in modeling users' dynamic interests.
    \item We develop a transition-aware embedding distillation module that distills the global item-to-item transition patterns into item embeddings to capture transitional signals, which serves as a calibration module for collaborative signals.
    \item We conduct extensive experiments on four real-world datasets to show the effectiveness of our proposed method. The results also highlight the different effects of the proposed two modules in improving recommendation performances.
\end{itemize}

\section{Preliminaries}

\subsection{Problem Formulation}
The sequential recommendation task aims to predict the next item that a user will interact with based on their historical interactions. Let $\mathcal{U}=\{u_1,u_2,\cdots,u_{|\mathcal{U}|}\}$ be the set of users, $\mathcal{I}=\{i_1,i_2,\cdots,i_{|\mathcal{I}|}\}$ be the set of items, and $S^{(u)}=[i_1^{(u)},i_2^{(u)},\cdots,i_{n_u}^{(u)}]$ be the interaction sequence of user $u$, where $n_u$ denotes the length of the sequence. The problem is formulated as calculating the probability that the next item will be interacted with, given the user's historical interactions:
\begin{equation}
    p\left(i_{n_u+1}^{(u)}|S^{(u)}\right).
\end{equation}
Then the top-N items will be recommended to user $u$ based on these probabilities in descending order.

\subsection{SASRec}
We first briefly introduce the \textit{SASRec} \cite{kang2018self} model, which is a state-of-the-art sequential recommender based on the self-attention module in \textit{Transformer} \cite{vaswani2017attention} and will be used as the base model in our approach. Given a user interaction sequence of the most recent $n$ items $[i_1,i_2,\cdots,i_n]$ (here we omit the superscript $(u)$ for simplicity), an embedding matrix $\mathbf{E}\in\mathbb{R}^{|\mathcal{I}|\times d}$ is used to convert the sequence into an embedding sequence $[\mathbf{e}_1,\mathbf{e}_2,\cdots,\mathbf{e}_n]$, where $d$ is the embedding size. Then a learnable positional embedding $\mathbf{P}\in\mathbb{R}^{n\times d}$ is added to encode the position information, resulting in $[\hat{\mathbf{e}}_1,\hat{\mathbf{e}}_2,\cdots,\hat{\mathbf{e}}_n]$, where $\hat{\mathbf{e}}_t=\mathbf{e}_t+\mathbf{p}_t$. Next, the transformer \cite{vaswani2017attention} module is used: 
\begin{equation}
    [\tilde{\mathbf{e}}_1,\tilde{\mathbf{e}}_2,\cdots,\tilde{\mathbf{e}}_n]=\textrm{Transformer}([\hat{\mathbf{e}}_1,\hat{\mathbf{e}}_2,\cdots,\hat{\mathbf{e}}_n]),
\end{equation}
 which adopts multiple blocks of self-attention and feed-forward networks. The self-attention layer is used to capture the long-term sequential dependency as follows:
\begin{equation}
    \label{equ:att}
    \textrm{Attention}(\mathbf{Q},\mathbf{K},\mathbf{V})=\mathrm{softmax}\left(\frac{\mathbf{Q}\mathbf{K}^T}{\sqrt{d}}\right)\mathbf{V},
\end{equation}
\begin{equation}
    \label{equ:qkv}
    \mathbf{Q}=\hat{\mathbf{E}}\mathbf{W}^Q,\ \mathbf{K}=\hat{\mathbf{E}}\mathbf{W}^K,\ \mathbf{V}=\hat{\mathbf{E}}\mathbf{W}^V,
\end{equation}
where $\mathbf{Q}$ represents the queries, $\mathbf{K}$ the keys, $\mathbf{V}$ the values, and $\mathbf{W}^Q$, $\mathbf{W}^K$, $\mathbf{W}^V\in\mathbb{R}^{d\times d}$ are the projection matrices for queries, keys, and values, respectively. Finally, the model predicts ranking scores by taking the dot product between the sequence embedding and the candidate item embeddings as $\hat{\mathbf{r}}_t=\tilde{\mathbf{e}}_t\mathbf{E}^T$. The cumulative cross-entropy loss is used for model training as follows:\footnote{This loss function has been shown more effective than the negative sampling-based binary cross-entropy loss \cite{li2023effective} and we use it for all models in our experiments.}
\begin{equation}
    \mathcal{L}_{rec}=-\sum_{t=1}^n \mathbf{r}_t\log \textrm{softmax}(\hat{\mathbf{r}}_t),
\end{equation}
where $\mathbf{r}_t\in\mathbb{R}^{1\times|\mathcal{I}|}$ is a one-hot vector converted from the index of the ground truth item at timestamp $t$.

\section{Methodology}

\begin{figure}[!t]
    \centering
    \includegraphics[width=3.35in]{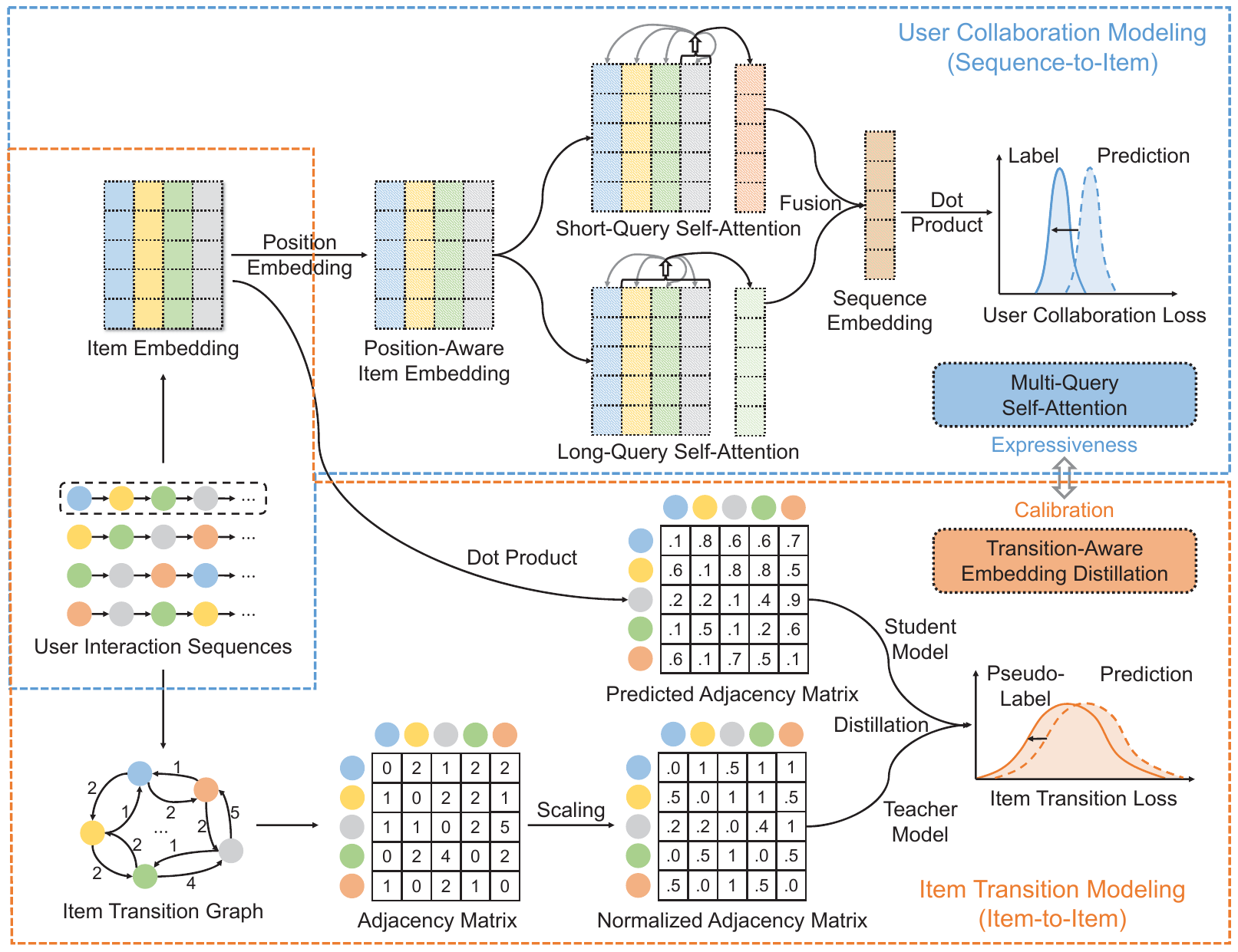}
    \caption{Illustration of the proposed MQSA-TED method. It consists of two main components: 1) Multi-Query Self-Attention for user collaboration modeling, and 2) Transition-Aware Embedding Distillation for item transition modeling.}
    \label{fig:framework}
\end{figure}

In this section, we present the proposed method, which consists of two main components as illustrated in Figure~\ref{fig:framework}: 1) Multi-Query Self-Attention for user collaboration modeling, and 2) Transition-Aware Embedding Distillation for item transition modeling.

\subsection{Multi-Query Self-Attention for User Collaboration Modeling}
We adopt SASRec as our base model owing to its strong ability to capture long-term sequential dependency and its state-of-the-art performance in sequential recommendation tasks \cite{kang2018self}. SASRec uses the self-attention module in Transformer \cite{vaswani2017attention}, whose main components are the \textit{queries}, \textit{keys}, and \textit{values}, as shown in Equation (\ref{equ:qkv}). Specifically, the attention query at timestamp $t$ in SASRec can be expressed as follows:
\begin{equation}
    \mathbf{q}_t=\hat{\mathbf{e}}_t\mathbf{W}^Q,
\end{equation}
where $\hat{\mathbf{e}}_t$ is the embedding vector of the item at timestamp $i$ after adding the positional embedding, and $\mathbf{W}^Q$ is a learnable projection matrix. Then, the attention weights assigned to historical items $[i_1,i_2,\cdots,i_{t}]$ at timestamp $t$ are determined by the scaled dot-product between the query embedding and the key embeddings as shown in Equation (\ref{equ:att}). Therefore, the attention weights are dominated by the single item at timestamp $t$, leading to a type of short-query self-attention.

However, this type of self-attention is limited in leveraging collaborative signals, especially when the item at timestamp $t$ is inconsistent with the user's primary preference. Specifically, SASRec can be viewed as a self-attention-enhanced first-order Markov chain model and its recommendation results can be significantly affected by a minor change in the order of users' interacted item sequences, such as swapping the position of the last two items. In other words, SASRec may generalize poorly on test samples lacking observed item transitions. However, real-world recommendation scenarios such as restaurant recommendations on Yelp have shown that user interests are relatively stable and less sensitive to the order of several recent choices \cite{zhu2021graph}, which SASRec may have difficulty in coping with. 
To address this limitation, we propose an $L$-query self-attention approach. First, we define the $L$-query self-attention as follows:
\begin{definition}[$L$-query Self-Attention]
An $L$-query self-attention is a type of self-attention module that uses the embeddings or their transformed representations of the most recent $L$ timestamps' items (tokens) as the attention query.
\end{definition}
Here we use the simple mean-pooling of the embeddings of the last $L$ items at timestamp $t$ as the query embedding:
\begin{equation}
    \tilde{\mathbf{q}}_t=\textrm{mean-pooling}(\hat{\mathbf{e}}_{t-L+1},\hat{\mathbf{e}}_{t-L+2},\cdots,\hat{\mathbf{e}}_{t})\tilde{\mathbf{W}}^{Q},
\end{equation}
where $L$ is a hyperparameter that controls the range of the attention query. Alternatively, other functions can be used to generate the query embedding, such as a weighted summation with time decay.

It is important to note that the hyperparameter $L$ controls the range of the historical context in self-attention. Using a large value of $L$ means that the model relies on long-range historical items to represent user interests, which contributes to capturing collaborative signals but may accumulate bias as user interests may shift over time. Conversely, using a small value of $L$ means that the model adopts the latest interacted items to represent user interests but can introduce variance due to the small number of used items. To balance the bias-variance trade-off, we propose a \textit{Multi-Query Self-Attention} (MQSA) method that combines the short-query self-attention (with $L=1$, similar to SASRec) with the long-query self-attention (with a larger $L$) using a hyperparameter $\alpha$:
\begin{equation}
    \tilde{\mathbf{e}}_t=\alpha\cdot\tilde{\mathbf{e}}_t^{short}+(1-\alpha)\cdot\tilde{\mathbf{e}}_t^{long}.
\end{equation}
Then, the sequence embedding $\tilde{\mathbf{e}}_t$ is used along with the embedding of candidate items to predict their ranking scores through dot product. Notably, we can also allow the model to learn the optimal $\alpha$. However, simultaneously learning the weights and the embeddings is challenging due to the inherent complexity. We could also incorporate more $L$s. We leave these for exploration in future work.

It is worth mentioning that the formulation of MQSA shares similar ideas with some approaches in the literature, such as FPMC \cite{rendle2010factorizing} and Fossil \cite{he2016fusing}, which explicitly model long-term user interests by employing user or item embeddings, respectively, and combine them with factorized Markov chains for sequential recommendation tasks. Compared to Fossil which uses the whole interacted items, MQSA introduces flexible window sizes of the last $L$ items to control the bias-variance trade-off. Furthermore, MQSA employs self-attention modules to enhance expressiveness, resulting in improved performance compared to the use of pure item embeddings in Fossil.

\subsection{Transition-Aware Embedding Distillation for Item Transition Modeling}

Sequential recommendation models have demonstrated their effectiveness in enhancing recommendation accuracy by capturing long-term user interests \cite{hidasi2015session,kang2018self,zhou2022filter}. However, these models may have limitations in leveraging the global item-to-item transitional signals. Specifically, most existing methods follow an auto-regressive framework \cite{kang2018self,zhou2022filter}. For each user, their preference at timestamp $t$ is learned based on their interacted items up to and including $t$, which is then used to predict the item at timestamp $t+1$. Nevertheless, this framework fails to enable the model to learn the global item-to-item transition patterns. In other words, the items not interacted with a user are treated equally, without considering the potential items that the current item $i_t$ is more likely to trigger.

To address this limitation, we propose a heuristic recommender based on item transitions and then develop a knowledge distillation method to integrate these global item transition patterns into sequential models. Specifically, we construct a global item transition graph $\mathcal{G}=(\mathcal{V},\mathcal{E})$ where $\mathcal{V}$ represents item nodes and $\mathcal{E}$ represents transition edges between items. $\mathcal{G}$ is a weighted and directed graph, where the weight of each edge represents the transition frequency between two items within a time span $k$, based on all user interaction sequences. Note that the time span hyperparameter $k$ is used to control the long-term item transition patterns and is set to $1$ by default (i.e. only transitions between directly adjacent items are considered). We use the adjacent matrix $\mathbf{A}\in\mathbb{R}^{|\mathcal{I}|\times|\mathcal{I}|}$ of $\mathcal{G}$ as the heuristic recommender, where $a_{i,j}$ is the transition frequency from item $i$ to item $j$, as shown in Figure~\ref{fig:framework}. It is a memory-based non-personalized method that recommends items based on the transition frequency from the current item to candidate items, as introduced in our preliminary experiments in Section 1. 

To distill the item transitions into the sequential model, we propose a \textit{Transition-Aware Embedding Distillation} (TED) method. First, we normalize the transition frequencies using a row normalization approach as $\bar{a}_{i,j}=\frac{a_{i,j}}{\max_j a_{i,j}}$. Then, we use a softmax function with temperature $\tau$ to generate pseudo-labels for knowledge distillation:
\begin{equation}
    \tilde{\mathbf{a}}_i=\textrm{softmax}(\bar{\mathbf{a}}_i/{\tau}),
\end{equation}
where a higher value of $\tau$ generates a softer probability distribution over items \cite{hinton2015distilling}. 

We adopt a simple factorization model as the student model, which predicts the item transition distribution of item $i$ by using the dot product between its embedding vector $\mathbf{e}_i$ and the embedding matrix $\mathbf{E}$ before the self-attention layers, where the dropout \cite{srivastava2014dropout} strategy is also used for robust learning. We apply the softmax function with temperature $\tau$ to obtain the predicted transition probabilities:
\begin{equation}
    \hat{\mathbf{a}}_i=\textrm{softmax}(\mathbf{e}_i\mathbf{E}^T/{\tau}).
\end{equation}
We use the cross-entropy loss to distill the item transitions into the sequential model by comparing the predicted and pseudo-label transition probabilities:
\begin{equation}
    \mathcal{L}_{kd}=-\sum_{i\in\mathcal{I}}\tilde{\mathbf{a}}_i\log\hat{\mathbf{a}}_i.
\end{equation}
Therefore, the factorization model can learn from the Item Transition model, enabling the item embeddings to memorize the item transition patterns. The overall loss function for the full model is:
\begin{equation}
    \mathcal{L}=\mathcal{L}_{rec}+\lambda_{kd}\mathcal{L}_{kd}+\lambda_{\Theta}||\Theta||_2^2,
\end{equation}
where $\Theta$ is the parameters, $\lambda_{kd}$ and $\lambda_{\Theta}$ are the hyperparameters that control the weights of distillation and $l_2$ regularization, respectively.

\subsection{Discussion}

\subsubsection{Relationship Between Two Modules}
Here we discuss the relationship between the user collaboration and item transition modules, and how they complement each other. 

\textbf{Expressiveness vs. Calibration.} The item transition module learns from a memory-based method that generates potential candidate items based on the global transition trends of the current item. However, it may generalize poorly to the items lacking observed transition patterns. On the other hand, the user collaboration module is a neural model that employs self-attentions to capture long-term user preferences and select the most likely next item based on historical items, resulting in a stronger ability to generalize but a limited ability to memorize and leverage item-to-item transition patterns. Therefore, the user collaboration model requires the item transition model to act as a calibrator for its predictions.

\textbf{Disentangled Learning.} The user collaboration and item transition modules are inherently disentangled, as we employ dual supervision where the original item embedding captures item-to-item transitional signals while the item embedding after self-attentions captures sequence-to-item collaborative signals. 

\textbf{Retrieval vs. Re-Ranking.} The item transition and user collaboration modules can be regarded as a retrieval model and a re-ranking model, respectively. The retrieval model provides insight into generating potential candidate items, while the re-ranking model provides insight into selecting the most relevant items for users based on their respective interaction histories.

\subsubsection{Comparison with Existing Methods} 

The proposed Transition-Aware Embedding Distillation (TED) module serves as a calibrator based on the item transition graph. Here we compare it with recent graph-based regularization methods:

\textbf{Graph Regularization (GraReg)} \cite{zhang2020graph} is a Euclidean distance-based regularization term on embedding layers using a $k$-nearest neighbor ($k$-NN) graph:
\begin{equation}
    \mathcal{L}=\mathcal{L}_{rec}+\lambda_{reg}\sum_{(i,j)\in \mathcal{E}}||\mathbf{e}_i-\mathbf{e}_j||^2,
\end{equation}
where $\lambda_{reg}$ is the coefficient hyperparameter for graph regularization, and $\mathcal{E}$ is the edges in the $k$-NN graph. We can use the transition frequency as the weights of the edges here. Therefore, GraReg uses the $k$ most related items for regularization, leading to learning localized transition patterns. Additionally, GraReg introduces an alignment loss but lacks a uniformity loss, where related items should be close to each other while unrelated ones should be separated \cite{wang2022towards}. In contrast, TED uses the global item transitions as the teacher model, enabling the item embeddings to memorize and leverage transitional signals.

\textbf{Graph-based Embedding Smoothing (GES)} \cite{zhu2021graph} employs graph convolutions on the global item transition graph for embedding smoothing in sequential recommenders:
\begin{equation}
    \mathbf{E}^{(l+1)}=\tilde{\mathbf{D}}^{-1/2}\tilde{\mathbf{A}}\tilde{\mathbf{D}}^{-1/2}\mathbf{E}^{(l)},
\end{equation}
where $\tilde{\mathbf{A}}=\mathbf{A}+\mathbf{I}$ is the adjacency matrix of the item transition graph with self-loops, $\tilde{\mathbf{D}}$ is the degree matrix of $\tilde{\mathbf{A}}$, and $l$ is the number of graph convolutional layers. However, stacking multiple graph convolutional layers may result in over-smoothing problems \cite{kipf2016semi}, potentially leading to a decline in model performance. In comparison, TED incorporates a hyperparameter to control the power of item transition distillation, allowing for flexibility in different recommendation scenarios.

\subsubsection{Model Complexity}
Here we analyze the space and time complexity of the proposed model. 

\textbf{Space Complexity.} The learnable parameters in SASRec are for item embeddings, positional embeddings, self-attention layers, feed-forward layers, and layer normalization. The total number of parameters in SASRec is $\mathcal{O}(|\mathcal{I}|d+nd+d^2)$ \cite{kang2018self}. Our proposed model introduces the long-query self-attention, which adds $\mathcal{O}(d^2)$ for projection matrices, feed-forward networks, and layer normalization. The embedding distillation module does not add any extra parameters. Therefore, the space complexity of our proposed model is the same as that of SASRec.

\textbf{Time Complexity.} The computational complexity of the self-attention layer and the feed-forward layer in SASRec is $\mathcal{O}(n^2d+nd^2)$. The cumulative cross-entropy loss has a complexity of $\mathcal{O}(|\mathcal{I}|nd)$. Thus, the total computational complexity of SASRec is $\mathcal{O}(|\mathcal{I}|nd+n^2d+nd^2)$. In our proposed model, the self-attention module has the same complexity as in SASRec. The embedding distillation module has a complexity of $\mathcal{O}(|\mathcal{I}|nd)$. Hence, the time complexity of the proposed model is the same as that of SASRec with the cumulative cross-entropy loss.

\section{Experiments}

We conduct experiments on four real-world datasets to evaluate the effectiveness of the proposed method.\footnote{The codes and datasets are available at https://github.com/zhuty16/MQSA-TED} The experiments are designed to answer the following research questions:

\begin{itemize}
\item [\textbf{RQ1.}] How does the proposed method compare with state-of-the-art sequential recommendation methods?
\item [\textbf{RQ2.}] How do the hyperparameters and various components affect the model performance?
\item [\textbf{RQ3.}] How does the proposed TED method compare with graph-based regularization methods?
\item [\textbf{RQ4.}] Can the proposed TED method benefit various recommendation models?
\item [\textbf{RQ5.}] How do the proposed two modules improve the model performance?

\end{itemize}

\subsection{Experimental Settings}

\subsubsection{Datasets}

\begin{table}[!t]
    \small
    \centering
    \caption{Summary of evaluation datasets. The datasets are from \cite{zhou2022filter}.}
    \begin{tabular}{lrrrrr}
        \hline
        Dataset & \# Users & \# Items & \# Actions & Density & Avg. Len. \\
        \hline
        Beauty & 22,363 & 12,101 & 198,502 & 0.073\% & 8.88 \\
        Sports & 25,598 & 18,357 & 296,337 & 0.063\% & 8.32 \\
        Toys & 19,412 & 11,924 & 167,597 & 0.072\% & 8.63 \\
        Yelp & 30,431 & 20,033 & 316,354 & 0.052\% & 10.40 \\
        \hline
    \end{tabular}
    \label{tab:dataset}
\end{table}

We adopt four datasets from \cite{zhou2022filter} for experiments. The Beauty, Sports, and Toys datasets are from the Amazon Review Dataset in \cite{mcauley2015image,he2016ups}.\footnote{https://cseweb.ucsd.edu/\textasciitilde jmcauley/datasets.html} The Yelp dataset is from the Yelp Open Dataset.\footnote{https://www.yelp.com/dataset} The training data, validation data, and test data are identical to those used in \cite{zhou2022filter}, which follows the leave-one-out evaluation protocol that treats the last item as the test data, the second last item as the validation data, and the remaining items as the training data for each user \cite{kang2018self}. The dataset statistics are shown in Table~\ref{tab:dataset}.

\begin{table*}[!t]
    \small
    \centering
    \caption{Performance comparison of different methods on four datasets. The best results are in boldface and the second best are underlined. Asterisk (*) indicates statistically significant improvements over the best baseline determined by a two-sample t-test ($p<0.01$) after repeating the experiments five times.}
    \begin{tabular}{llcccccccccc}
        \hline
        Dataset & Metric & POP & LightGCN & FPMC & Caser & GRU4Rec & SASRec & BERT4Rec & FMLP-Rec & MQSA-TED & Improv. \\
        \hline
        \multirow{6}{*}{Beauty} & HR@5 & 0.0077 & 0.0374 & 0.0596 & 0.0359 & 0.0489 & 0.0694 & 0.0419 & \underline{0.0698} & \textbf{0.0752}* & 7.23\% \\
        & NDCG@5 & 0.0042 & 0.0247 & 0.0419 & 0.0241 & 0.0342 & \underline{0.0492} & 0.0275 & 0.0488 & \textbf{0.0534}* & 8.58\% \\
        & HR@10 & 0.0135 & 0.0571 & 0.0838 & 0.0511 & 0.0695 & 0.0932 & 0.0647 & \underline{0.0995} & \textbf{0.1039}* & 4.44\% \\
        & NDCG@10 & 0.0061 & 0.0311 & 0.0497 & 0.0290 & 0.0408 & 0.0568 & 0.0349 & \underline{0.0583} & \textbf{0.0627}* & 7.48\% \\
        & HR@20 & 0.0217 & 0.0841 & 0.1151 & 0.0720 & 0.0998 & 0.1286 & 0.0992 & \underline{0.1361} & \textbf{0.1435}* & 5.40\% \\
        & NDCG@20 & 0.0081 & 0.0379 & 0.0576 & 0.0343 & 0.0484 & 0.0657 & 0.0435 & \underline{0.0675} & \textbf{0.0726}* & 7.62\% \\
        \hline
        \multirow{6}{*}{Sports} & HR@5 & 0.0057 & 0.0252 & 0.0337 & 0.0195 & 0.0221 & 0.0380 & 0.0241 & \underline{0.0415} & \textbf{0.0455}* & 9.52\% \\
        & NDCG@5 & 0.0041 & 0.0170 & 0.0234 & 0.0128 & 0.0143 & 0.0267 & 0.0161 & \underline{0.0287} & \textbf{0.0320}* & 11.34\% \\
        & HR@10 & 0.0091 & 0.0384 & 0.0499 & 0.0290 & 0.0357 & 0.0541 & 0.0380 & \underline{0.0598} & \textbf{0.0643}* & 7.48\% \\
        & NDCG@10 & 0.0052 & 0.0212 & 0.0286 & 0.0159 & 0.0187 & 0.0318 & 0.0206 & \underline{0.0346} & \textbf{0.0380}* & 9.85\% \\
        & HR@20 & 0.0175 & 0.0576 & 0.0703 & 0.0431 & 0.0548 & 0.0752 & 0.0583 & \underline{0.0847} & \textbf{0.0906}* & 6.93\% \\
        & NDCG@20 & 0.0073 & 0.0260 & 0.0337 & 0.0195 & 0.0235 & 0.0371 & 0.0257 & \underline{0.0409} & \textbf{0.0446}* & 9.09\% \\
        \hline
        \multirow{6}{*}{Toys} & HR@5 & 0.0065 & 0.0378 & 0.0664 & 0.0307 & 0.0420 & 0.0736 & 0.0379 & \underline{0.0785} & \textbf{0.0834}* & 6.24\% \\
        & NDCG@5 & 0.0044 & 0.0251 & 0.0463 & 0.0224 & 0.0297 & 0.0533 & 0.0244 & \underline{0.0570} & \textbf{0.0600}* & 5.31\% \\
        & HR@10 & 0.0090 & 0.0564 & 0.0925 & 0.0420 & 0.0597 & 0.0989 & 0.0589 & \underline{0.1062} & \textbf{0.1130}* & 6.42\% \\
        & NDCG@10 & 0.0052 & 0.0311 & 0.0547 & 0.0260 & 0.0354 & 0.0615 & 0.0312 & \underline{0.0659} & \textbf{0.0696}* & 5.56\% \\
        & HR@20 & 0.0143 & 0.0795 & 0.1212 & 0.0597 & 0.0834 & 0.1299 & 0.0857 & \underline{0.1399} & \textbf{0.1503}* & 7.41\% \\
        & NDCG@20 & 0.0065 & 0.0370 & 0.0619 & 0.0305 & 0.0414 & 0.0693 & 0.0379 & \underline{0.0743} & \textbf{0.0789}* & 6.23\% \\
        \hline
        \multirow{6}{*}{Yelp} & HR@5 & 0.0056 & \underline{0.0290} & 0.0272 & 0.0199 & 0.0211 & 0.0232 & 0.0264 & 0.0270 & \textbf{0.0320}* & 10.18\% \\
        & NDCG@5 & 0.0036 & \underline{0.0184} & 0.0173 & 0.0129 & 0.0134 & 0.0151 & 0.0169 & 0.0169 & \textbf{0.0205}* & 11.74\% \\
        & HR@10 & 0.0096 & \underline{0.0486} & 0.0433 & 0.0334 & 0.0367 & 0.0379 & 0.0441 & 0.0446 & \textbf{0.0517}* & 6.36\% \\
        & NDCG@10 & 0.0049 & \underline{0.0246} & 0.0224 & 0.0172 & 0.0184 & 0.0198 & 0.0226 & 0.0225 & \textbf{0.0269}* & 8.95\% \\
        & HR@20 & 0.0158 & \underline{0.0790} & 0.0695 & 0.0535 & 0.0603 & 0.0623 & 0.0737 & 0.0721 & \textbf{0.0832}* & 5.24\% \\
        & NDCG@20 & 0.0065 & \underline{0.0323} & 0.0290 & 0.0222 & 0.0244 & 0.0259 & 0.0300 & 0.0294 & \textbf{0.0348}* & 7.62\% \\
        \hline
    \end{tabular}
    \label{tab:main_result}
\end{table*}

\begin{figure}[!t]
    \centering
    \subfloat{\includegraphics[width=1.5in]{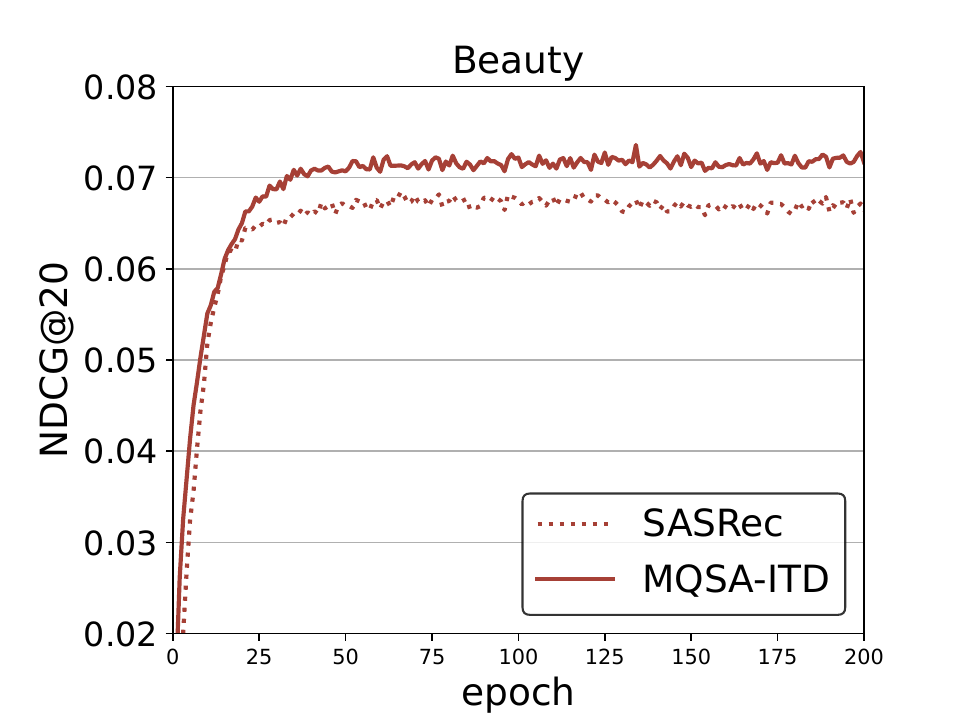}}
    \hfil
    \subfloat{\includegraphics[width=1.5in]{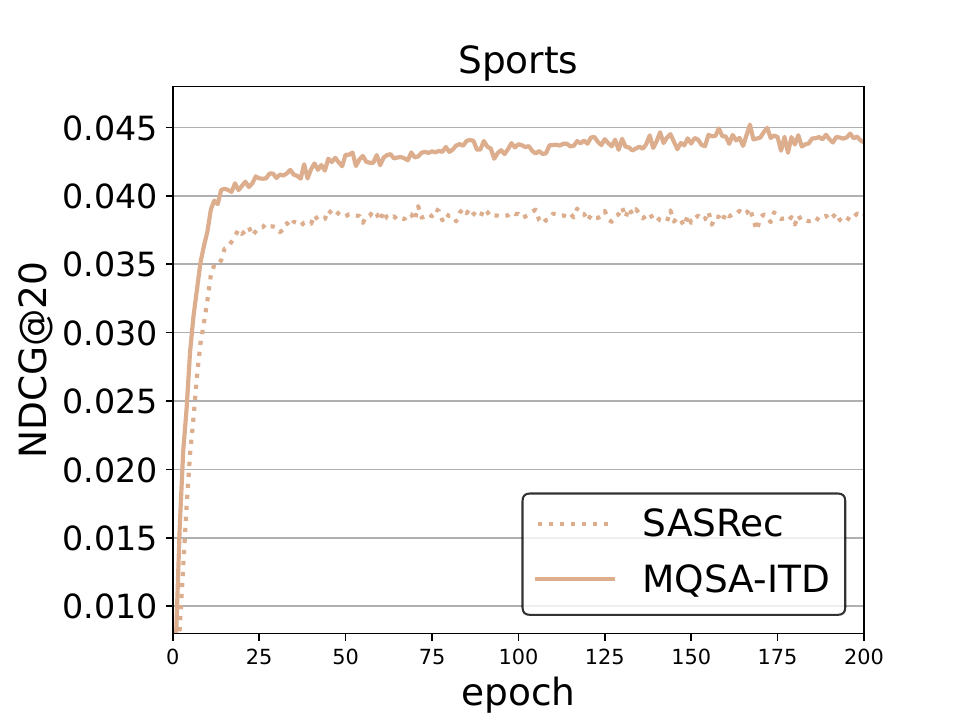}}
    \hfil
    \subfloat{\includegraphics[width=1.5in]{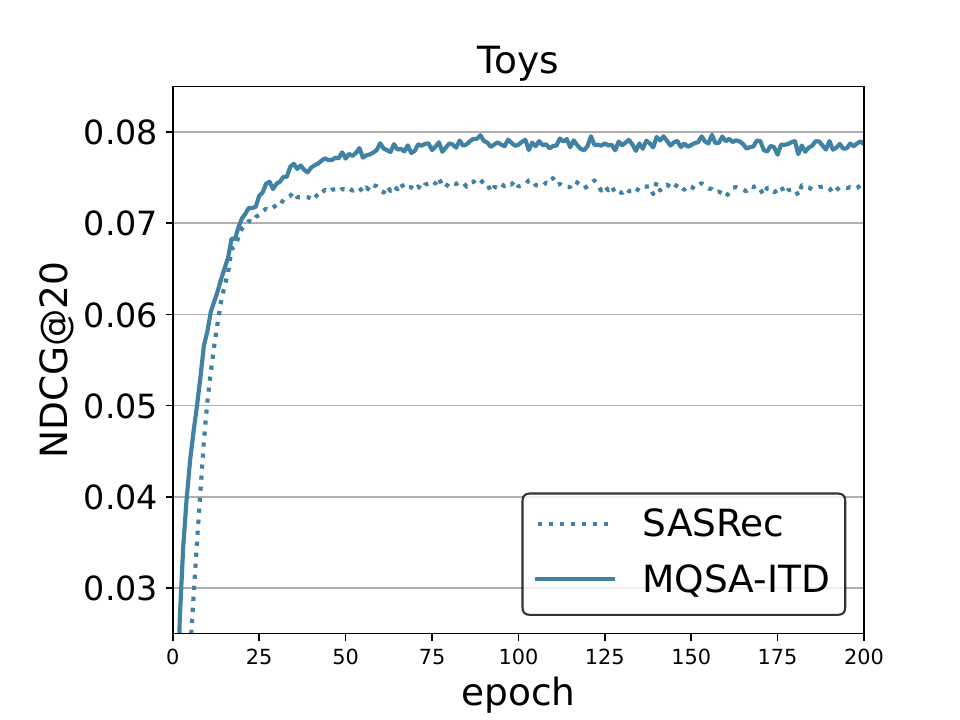}}
    \hfil
    \subfloat{\includegraphics[width=1.5in]{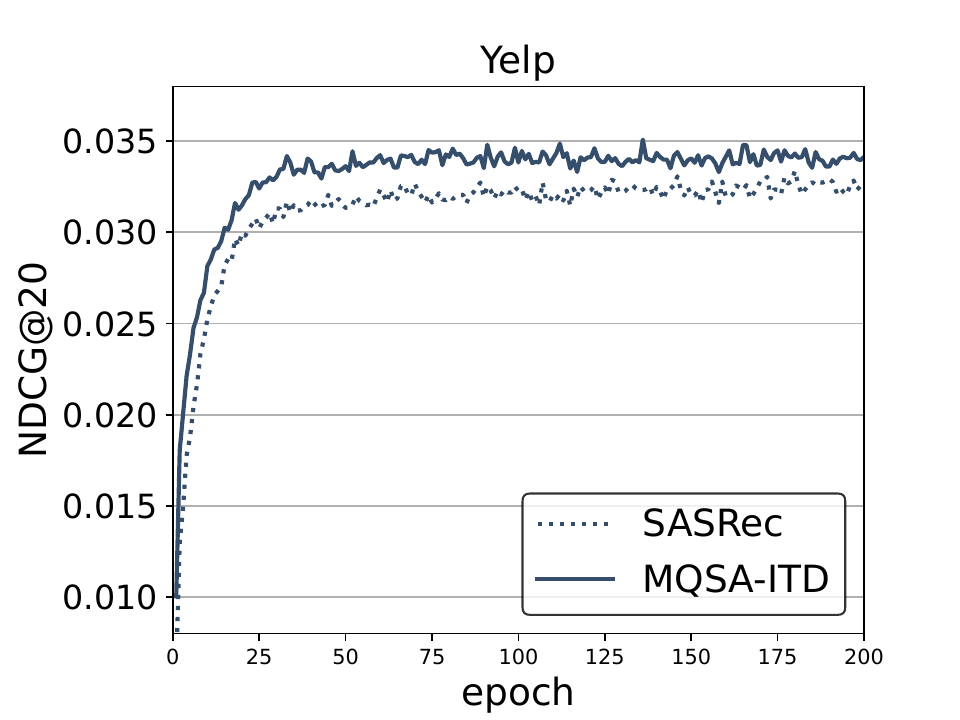}}
    \caption{Performance curves of SASRec and our proposed MQSA-TED on four datasets.}
    \label{fig:epoch}
\end{figure}

\subsubsection{Baselines}
We compare the proposed method with various types of state-of-the-art baselines in sequential recommendation:
\begin{itemize}
\item \textbf{POP}: a non-personalized method that ranks items based on their popularity.
\item \textbf{LightGCN} \cite{he2020lightgcn}: a \textit{GCN}-based method that learns user and item embeddings through linear propagation on the user-item interaction graph.
\item \textbf{FPMC} \cite{rendle2010factorizing}: a \textit{Markov chain}-based method that combines matrix factorization and factorized Markov chains.
\item \textbf{Caser} \cite{tang2018personalized}: a \textit{CNN}-based method that uses horizontal and vertical convolutions to learn sequential patterns.
\item \textbf{GRU4Rec} \cite{hidasi2015session}: an \textit{RNN}-based method that uses Gated Recurrent Units (GRU) to model dynamic user preferences.
\item \textbf{SASRec} \cite{kang2018self}: a \textit{unidirectional Transformer}-based method that models user interests using the self-attention module in Transformer \cite{vaswani2017attention}.
\item \textbf{BERT4Rec} \cite{sun2019bert4rec}: a \textit{bidirectional Transformer}-based method that models user interests using the self-attention module in BERT \cite{devlin2018bert}.
\item \textbf{FMLP-Rec} \cite{zhou2022filter}: an \textit{MLP}-based method that is currently the state-of-the-art sequential recommendation model based on filter-enhanced MLP.
\end{itemize}

\subsubsection{Evaluation Metrics and Protocols}
We adopt Hit Ratio@N (HR@N) and NDCG@N to evaluate the performance of the compared methods on the sequential recommendation task \cite{zhou2020s3,zhou2022filter}. We set $N$ to $5$, $10$, and $20$ by default and report the average scores of users. For each user, we rank all items except for the positive ones in their training or validation data \cite{krichene2022sampled}. To ensure the robustness of the results, we randomly initialize each model five times and report the average performance.

\subsubsection{Implementation and Hyperparameter Settings}
We implement all models with TensorFlow and use the cross-entropy loss for all models for a fair comparison, which has been proved to outperform the negative sampling-based losses significantly \cite{li2023effective}. For common hyperparameters in all models, the maximum sequence length is set to $50$, the embedding size $d$ is set to $64$, the learning rate is tuned in \{5e-3, 1e-3, 5e-4, 1e-4\}, and the $l_2$ regularization is tuned in \{0, 1e-6, 1e-5, 1e-4, 1e-3\}. All models are trained with mini-batch Adam \cite{kingma2014adam} and the batch size is set to $256$. Other hyperparameters of different models are tuned on the validation set according to the suggestions in their respective papers. The results of baseline methods under their optimal hyperparameter settings are reported.

\subsection{Main Results (RQ1)}
Table~\ref{tab:main_result} presents a performance comparison of different methods. The results show that, on Amazon datasets, sequential methods such as FPMC, SASRec, and FMLP-Rec outperform the non-sequential method LightGCN significantly. Among the sequential methods, FMLP-Rec performs the best. However, on the Yelp dataset, LightGCN outperforms the sequential methods due to the weak sequentiality of user interactions on Yelp \cite{zhu2021graph}. Furthermore, our proposed method significantly outperforms all baseline methods, with an average improvement of $6.24\%$ in Hit Ratio@20 and $7.64\%$ in NDCG@20 compared to the best baseline.

Figure~\ref{fig:epoch} shows the performances of SASRec and our proposed method with respect to the training epochs. One can observe that our proposed method consistently outperforms SASRec by a notable margin, showing the effectiveness of the proposed modules.

\begin{figure}[!t]
    \centering
    \subfloat{\includegraphics[width=0.83in]{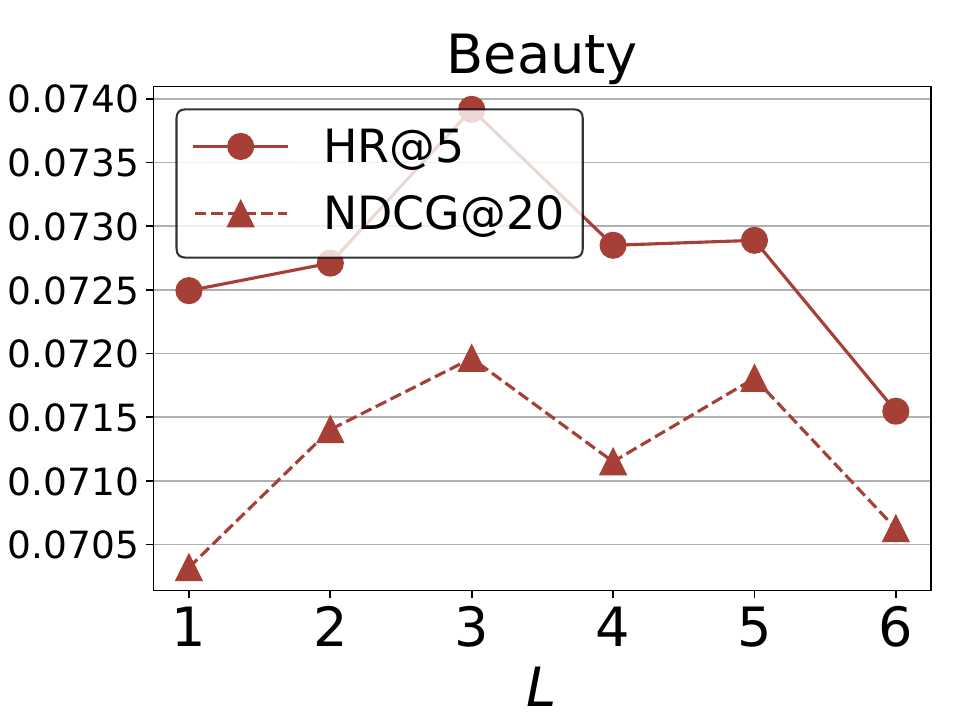}}
    \hfil
    \subfloat{\includegraphics[width=0.83in]{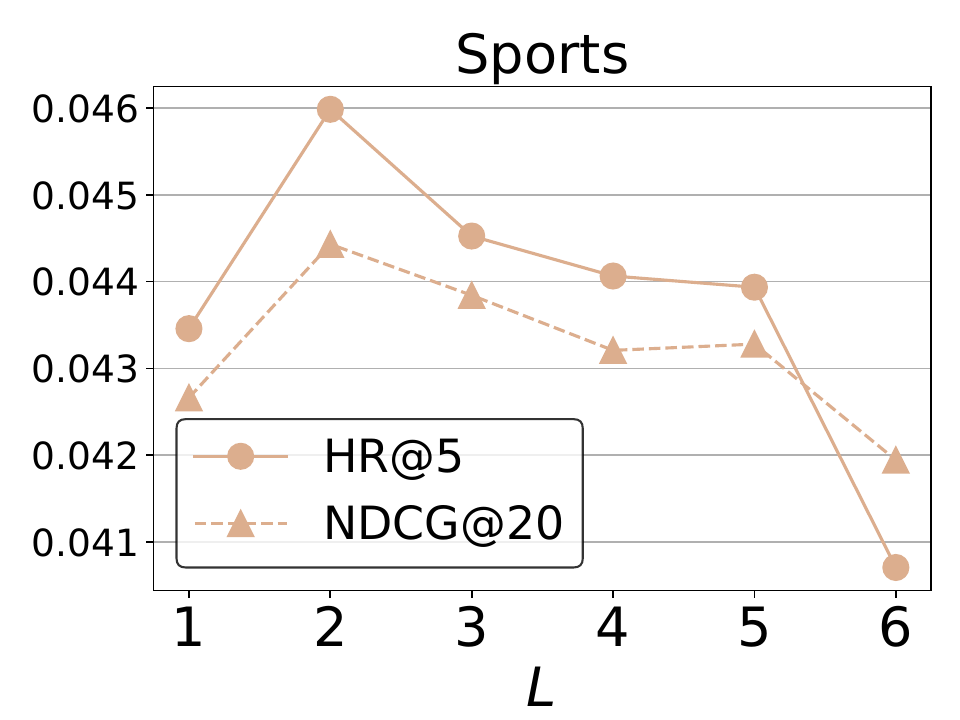}}
    \hfil
    \subfloat{\includegraphics[width=0.83in]{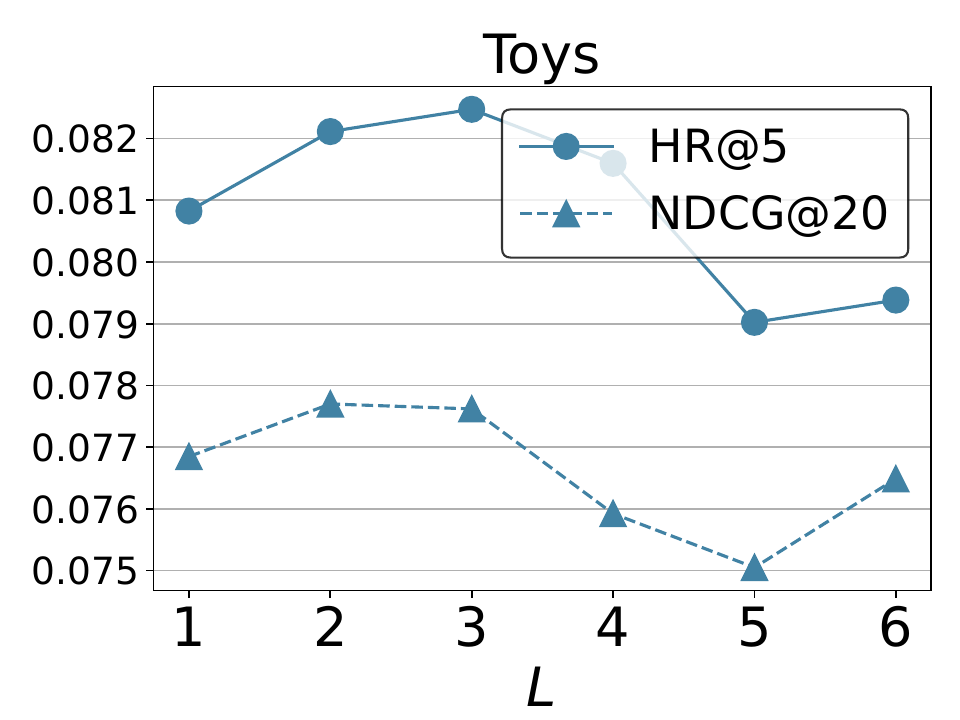}}
    \hfil
    \subfloat{\includegraphics[width=0.83in]{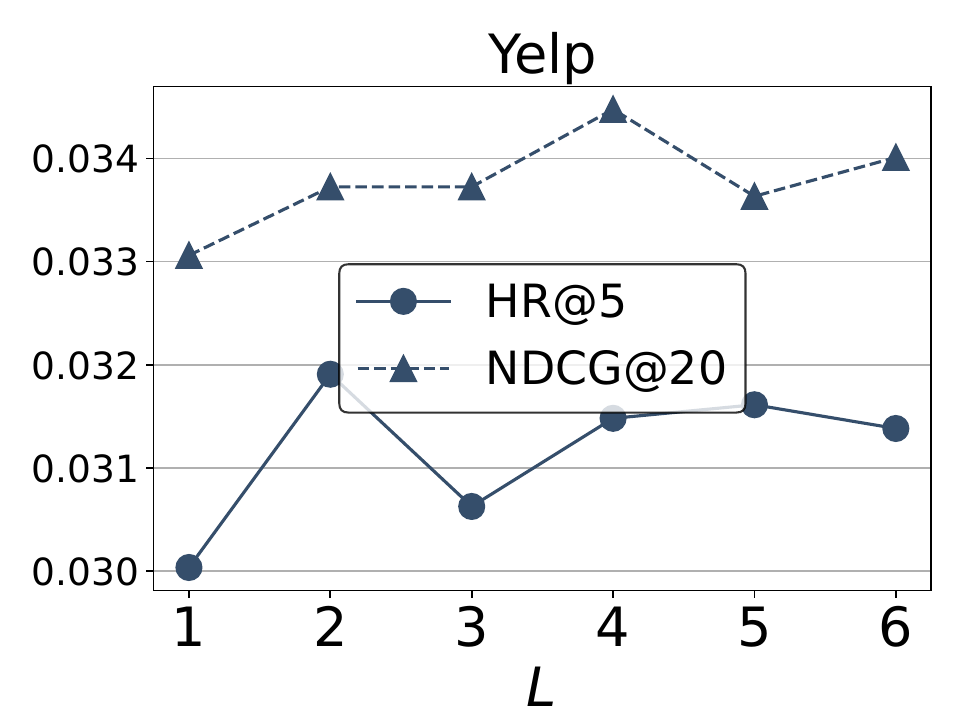}}
    
    \hfil
    \subfloat{\includegraphics[width=0.83in]{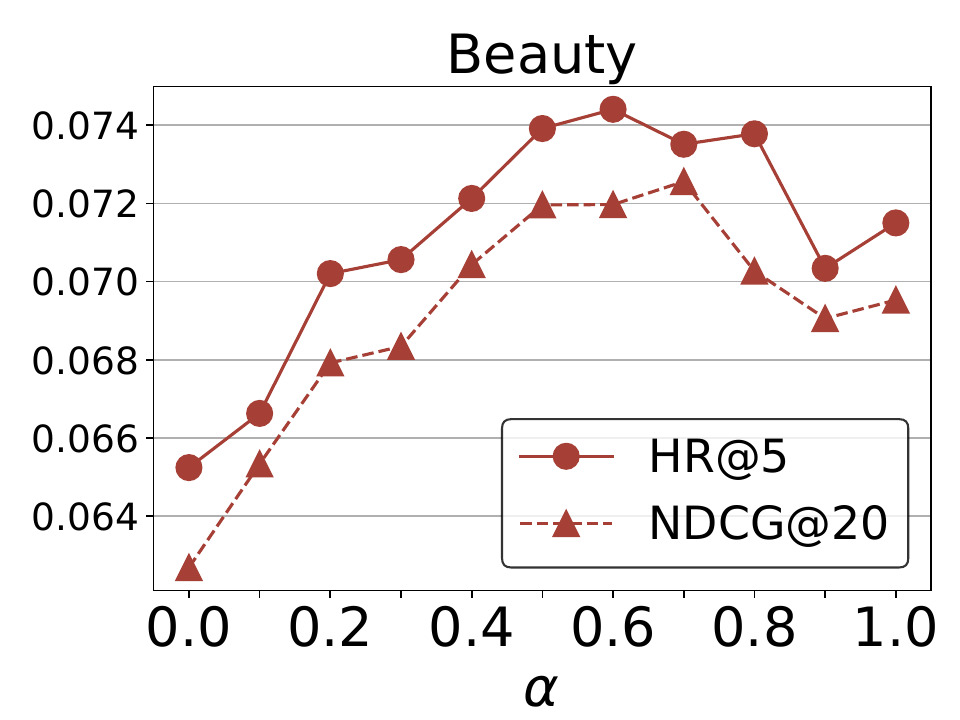}}
    \hfil
    \subfloat{\includegraphics[width=0.83in]{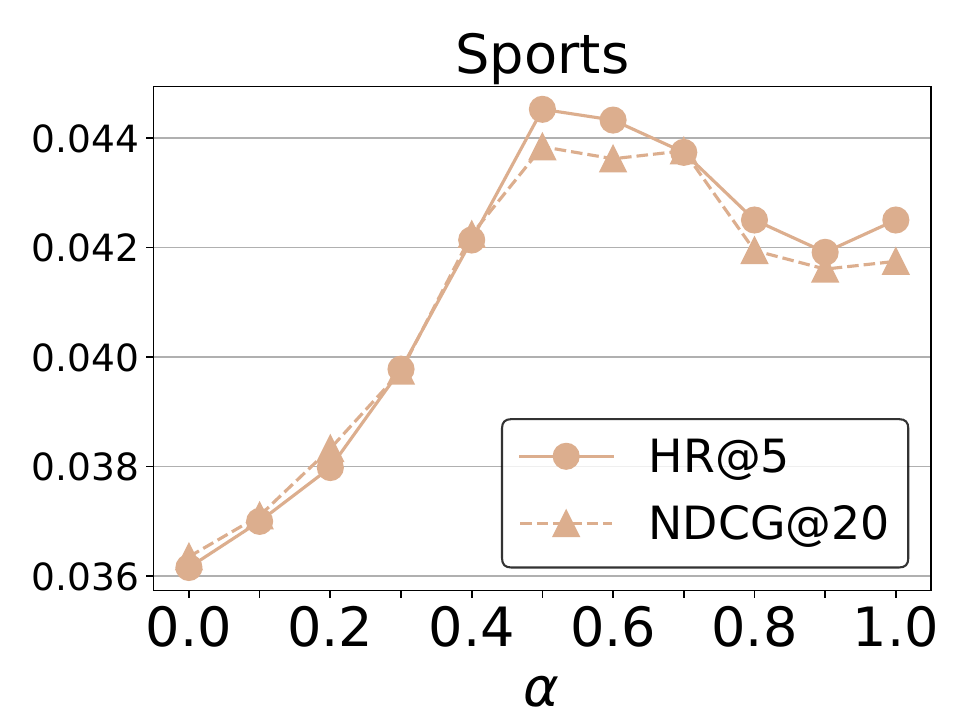}}
    \hfil
    \subfloat{\includegraphics[width=0.83in]{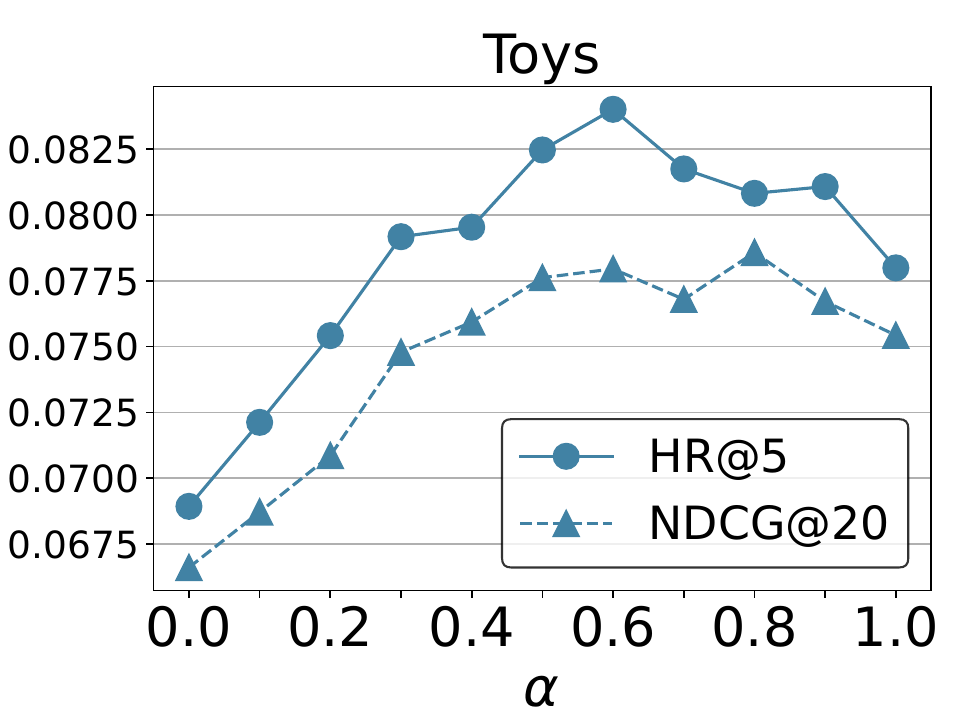}}
    \hfil
    \subfloat{\includegraphics[width=0.83in]{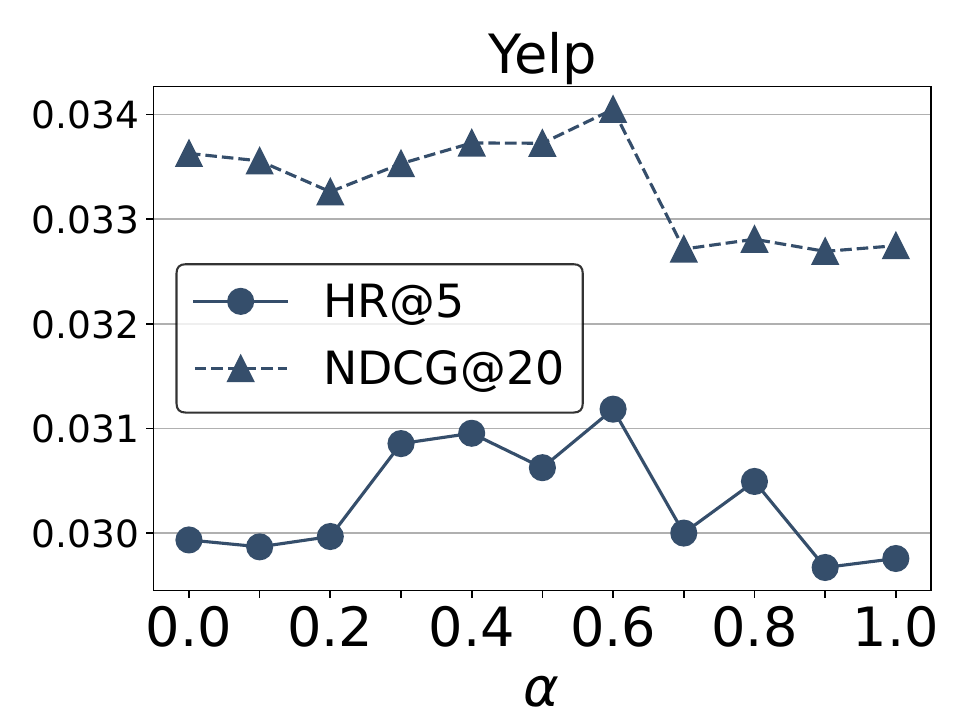}}

    \hfil
    \subfloat{\includegraphics[width=0.83in]{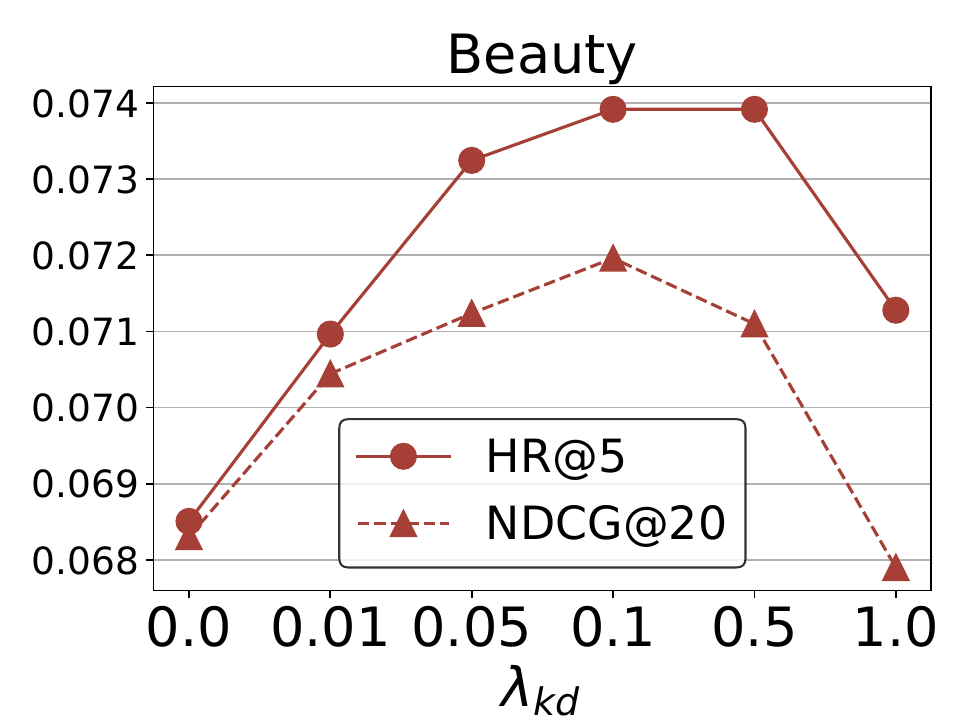}}
    \hfil
    \subfloat{\includegraphics[width=0.83in]{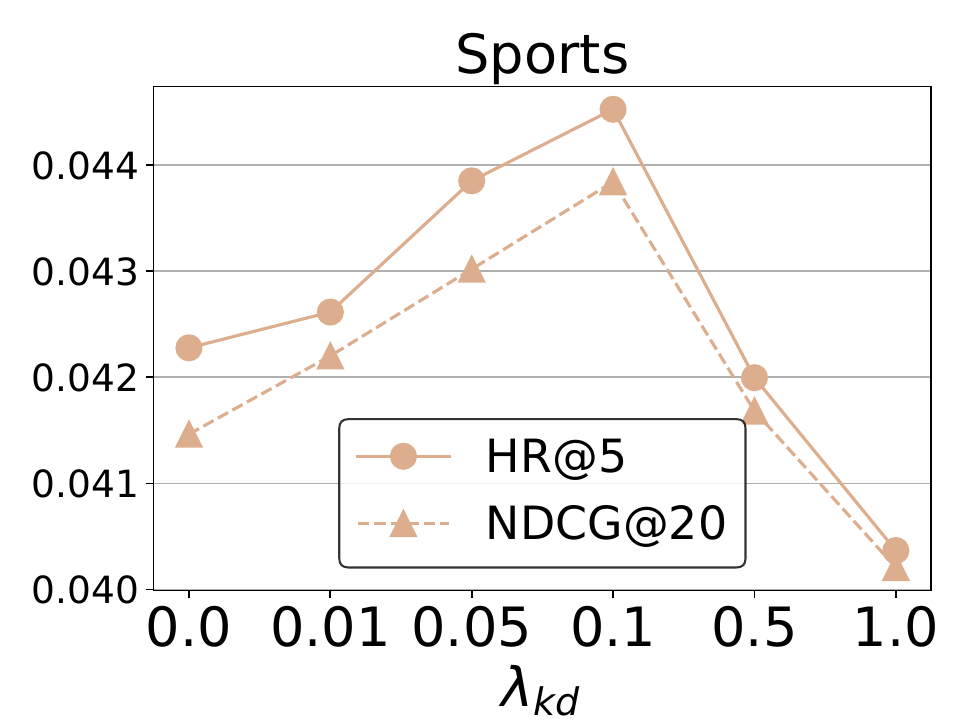}}
    \hfil
    \subfloat{\includegraphics[width=0.83in]{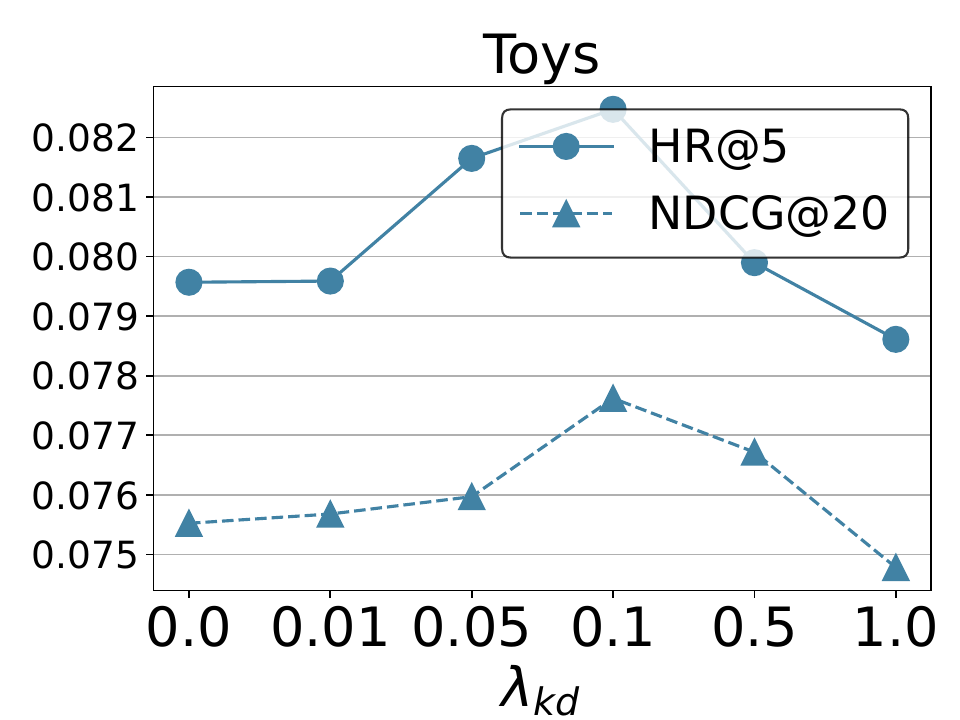}}
    \hfil
    \subfloat{\includegraphics[width=0.83in]{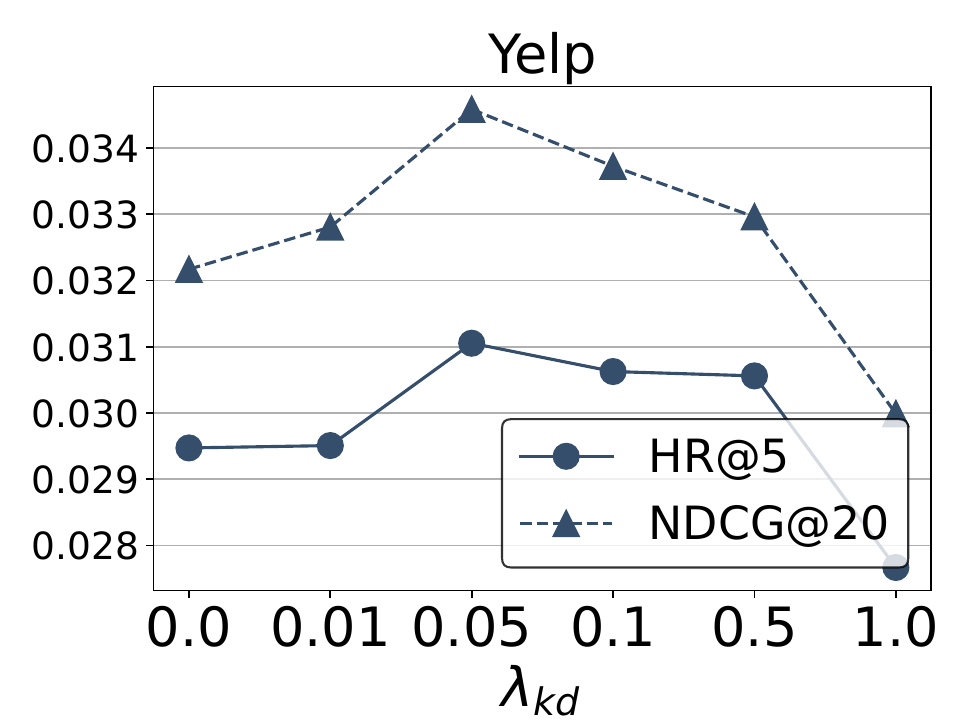}}

    \hfil
    \subfloat{\includegraphics[width=0.83in]{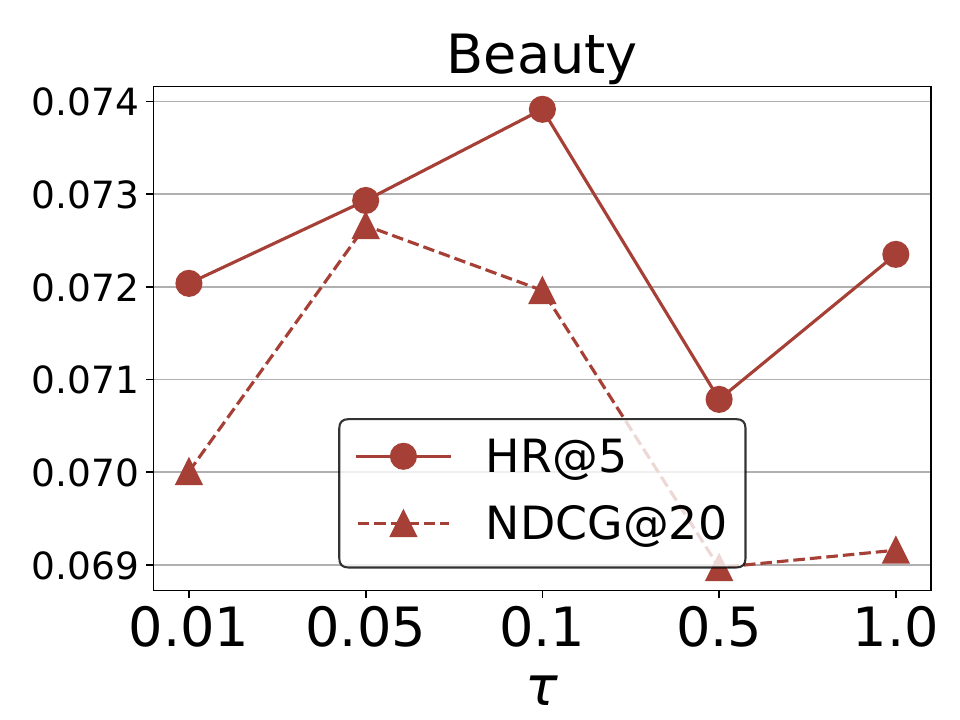}}
    \hfil
    \subfloat{\includegraphics[width=0.83in]{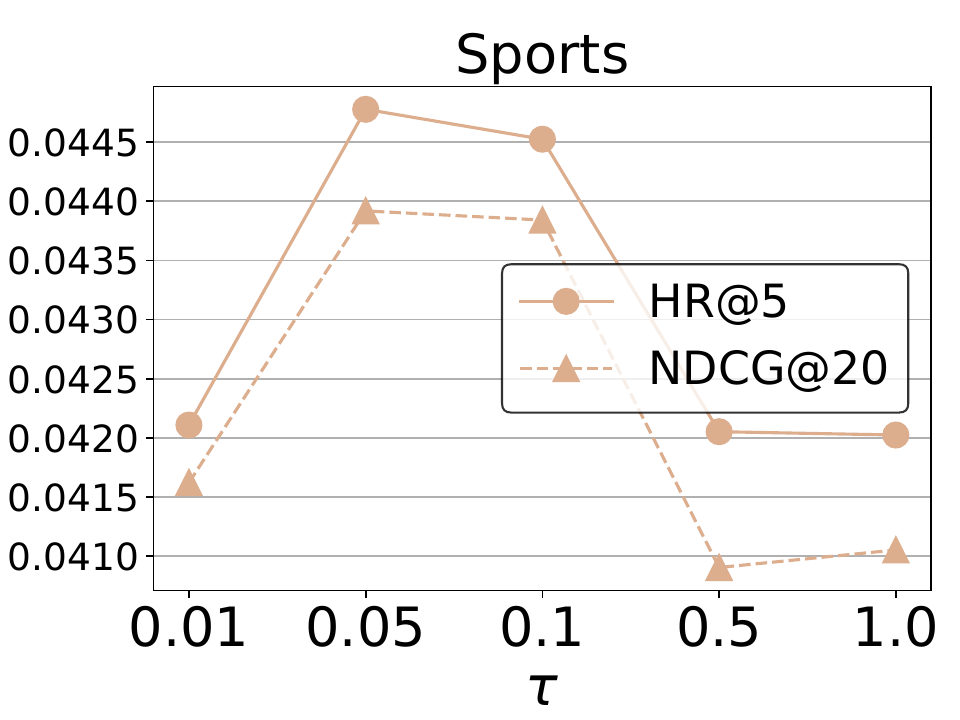}}
    \hfil
    \subfloat{\includegraphics[width=0.83in]{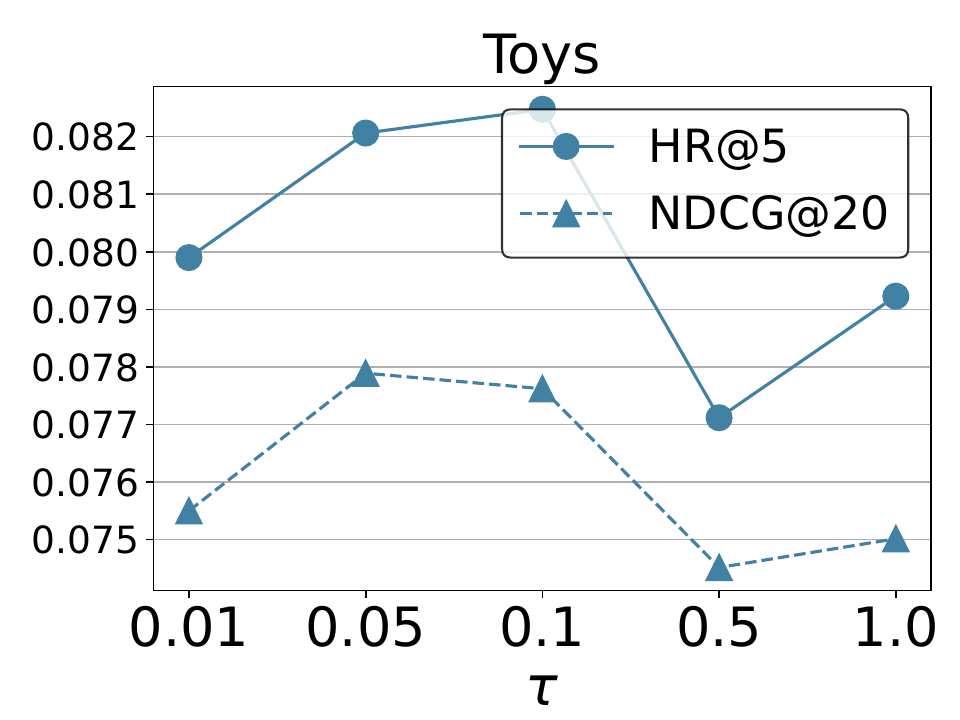}}
    \hfil
    \subfloat{\includegraphics[width=0.83in]{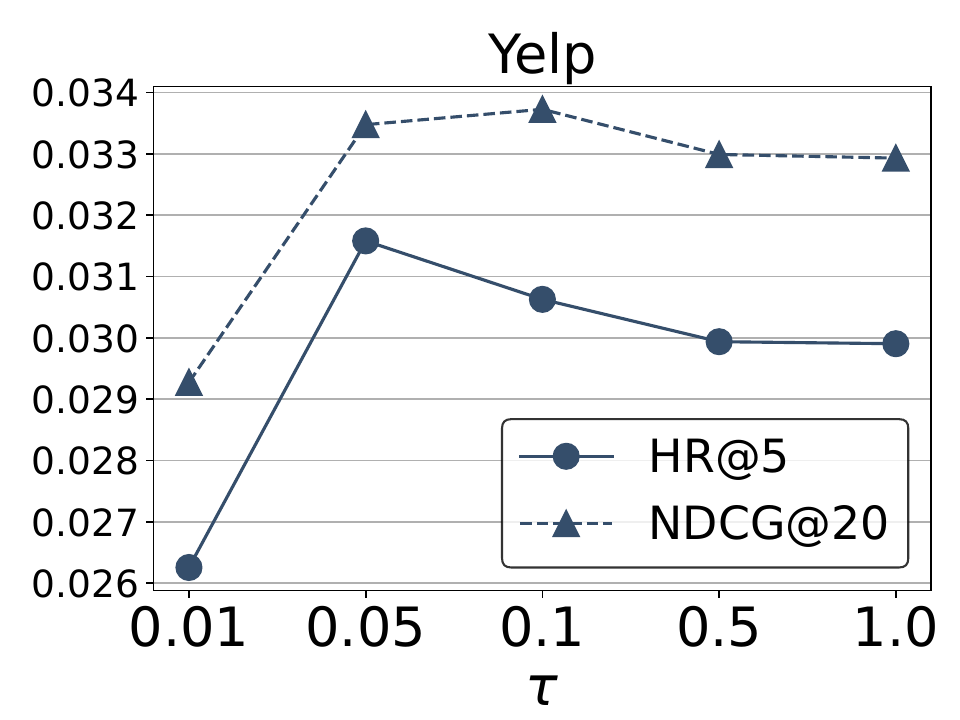}}
    \caption{Performance of the proposed MQSA-TED w.r.t. various hyperparameters on four datasets.}
    \label{fig:hyperparameter}
\end{figure}

\subsection{Hyperparameter and Ablation Studies (RQ2)}

Figure~\ref{fig:hyperparameter} presents the performance of our proposed method with respect to various hyperparameters and modules:

\subsubsection{Length of Long-Query Self-Attention $L$.} It can be observed the best $L$ depends on the datasets and the model generally performs well when $L$ is in the range of $[2,4]$, showing the effectiveness of long-query self-attention in capturing collaborative signals.

\subsubsection{Balance of Long and Short-Query Self-Attention $\alpha$.} The results show that when $\alpha$ is approximately $0.5$, the model achieves the best performance, indicating a proper bias-variance trade-off in modeling user interests. Notably, when $\alpha=1$, the model degrades to SASRec with TED. Therefore, the proposed multi-query self-attention significantly outperforms the short-query self-attention used in SASRec with a proper $\alpha$.

\subsubsection{Weight of Embedding Distillation $\lambda_{kd}$.} It can be seen that the model performs better when $\lambda_{kd}$ is approximately $0.1$, demonstrating the effectiveness of the TED module. Note that when $\lambda_{kd}=0$, our proposed method degrades to the MQSA model without TED, resulting in a significant drop in performance.

\subsubsection{Temperature of Embedding Distillation $\tau$.} The results suggest that the model requires relatively hard pseudo-labels of item transition distributions for effective knowledge distillation, as the best performance is achieved when $\tau=0.05$ or $\tau=0.1$.

\begin{table}[!t]
    \small
    \centering
    \caption{Performance comparison of the proposed TED module with graph-based methods on four datasets. The best results are in boldface and the second best are underlined.}
    \begin{tabular}{llcccc}
        \hline
        Dataset & Metric & MQSA & +GES & +GraReg & +TED \\
        \hline
        \multirow{2}{*}{Beauty} 
        & NDCG@10 & 0.0599 & \underline{0.0623} & 0.0611 & \textbf{0.0627} \\
        & NDCG@20 & 0.0694 & \underline{0.0724} & 0.0708 & \textbf{0.0726} \\
        \hline
        \multirow{2}{*}{Sports} 
        & NDCG@10 & 0.0344 & \underline{0.0370} & 0.0351 & \textbf{0.0380} \\
        & NDCG@20 & 0.0408 & \underline{0.0434} & 0.0416 & \textbf{0.0446} \\
        \hline
        \multirow{2}{*}{Toys} 
        & NDCG@10 & 0.0654 & \underline{0.0672} & 0.0667 & \textbf{0.0696} \\
        & NDCG@20 & 0.0749 & \underline{0.0765} & 0.0755 & \textbf{0.0789} \\
        \hline
        \multirow{2}{*}{Yelp} 
        & NDCG@10 & 0.0255 & 0.0244 & \underline{0.0257} & \textbf{0.0269} \\
        & NDCG@20 & 0.0327 & 0.0320 & \underline{0.0330} & \textbf{0.0348} \\
        \hline
    \end{tabular}
    \label{tab:ablation}
\end{table}

\begin{table}[!t]
    \small
    \centering
    \caption{Performance comparison of LightGCN and FMLP-Rec w/ and w/o the proposed TED module on four datasets. The best results under each backbone are in boldface.}
    \begin{tabular}{llcccc}
        \hline
        Dataset & Metric & LightGCN & +TED & FMLP-Rec & +TED \\
        \hline
        \multirow{2}{*}{Beauty} 
            & NDCG@10 & 0.0311 & \textbf{0.0399} & 0.0583 & \textbf{0.0596} \\
            & NDCG@20 & 0.0379 & \textbf{0.0484} & 0.0675 & \textbf{0.0684} \\
        \hline
        \multirow{2}{*}{Sports}
            & NDCG@10 & 0.0212 & \textbf{0.0246} & 0.0346 & \textbf{0.0356} \\
            & NDCG@20 & 0.0260 & \textbf{0.0298} & 0.0409 & \textbf{0.0423} \\
        \hline
        \multirow{2}{*}{Toys} 
            & NDCG@10 & 0.0311 & \textbf{0.0388} & 0.0659 & \textbf{0.0675} \\
            & NDCG@20 & 0.0370 & \textbf{0.0459} & 0.0743 & \textbf{0.0762} \\
        \hline
        \multirow{2}{*}{Yelp} 
            & NDCG@10 & \textbf{0.0246} & 0.0236 & 0.0225 & \textbf{0.0226} \\
            & NDCG@20 & \textbf{0.0323} & 0.0312 & 0.0294 & \textbf{0.0296} \\
        \hline
    \end{tabular}
    \label{tab:ITD}
\end{table}

\begin{figure}[!t]
    \centering
    \subfloat{\includegraphics[width=1.5in]{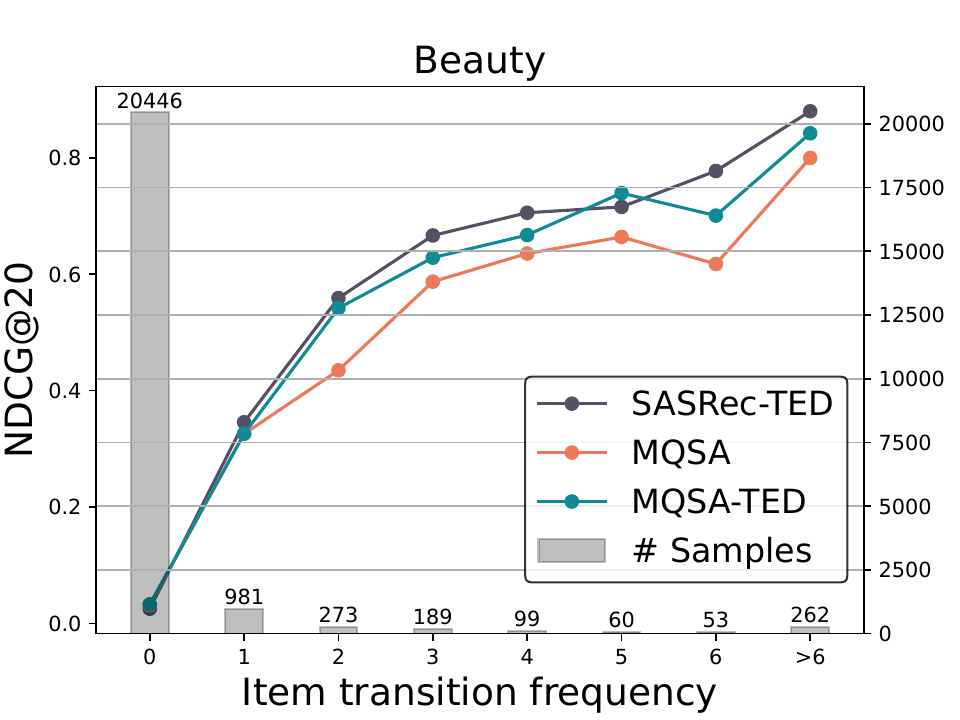}}
    \hfil
    \subfloat{\includegraphics[width=1.5in]{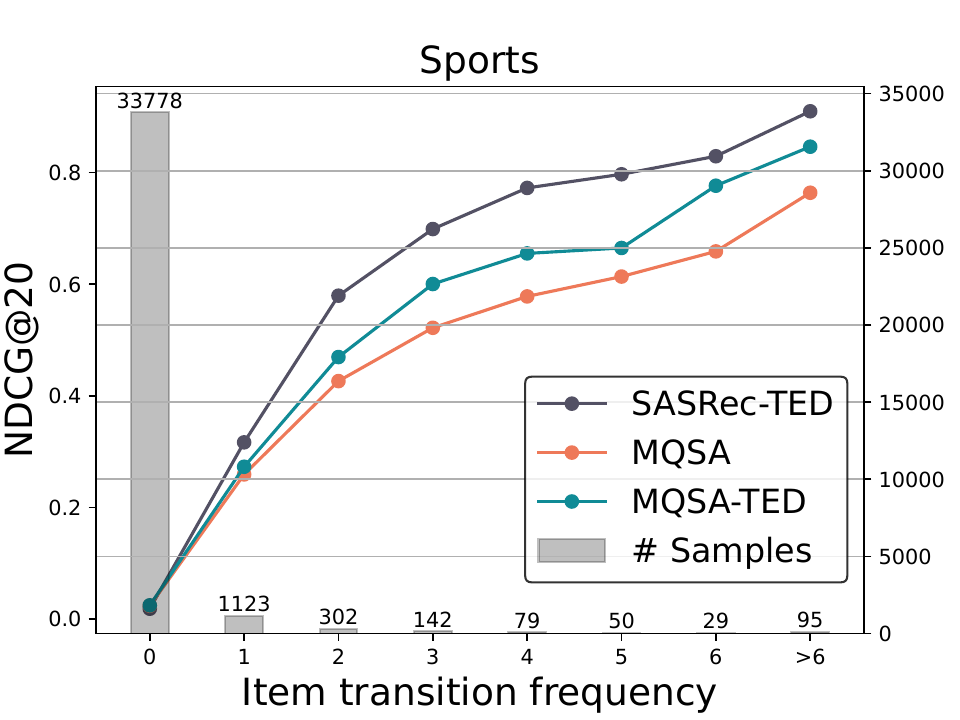}}
    \hfil
    \subfloat{\includegraphics[width=3.0in]{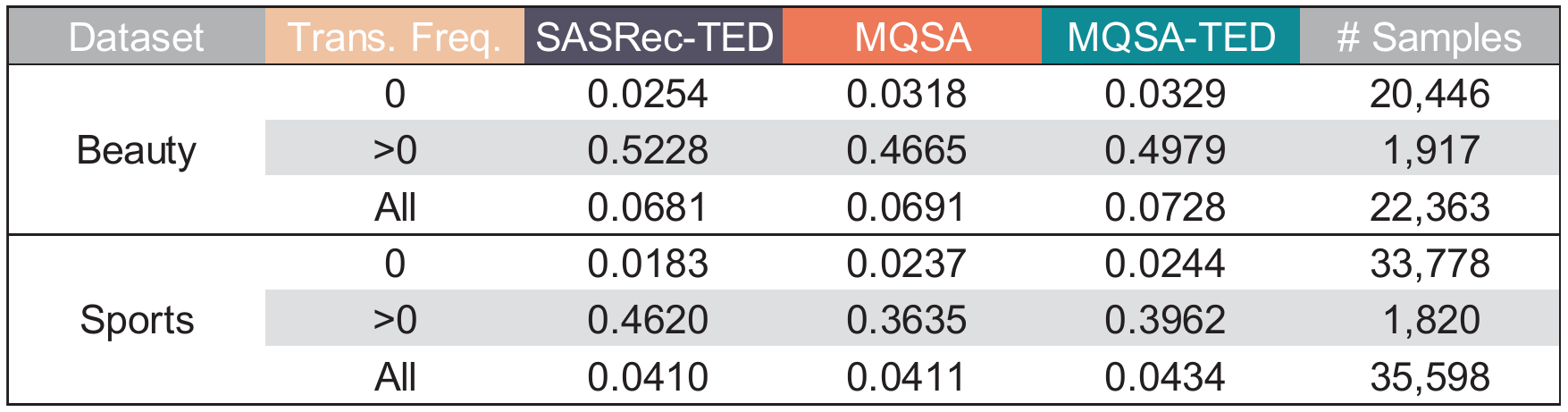}}
    \caption{Performance of three methods w.r.t. item transition frequency on two datasets. MQSA-TED outperforms MQSA on test samples with frequent transitions and outperforms SASRec-TED on test samples lacking transition instances.}
    \label{fig:group_exper}
\end{figure}

\subsection{Comparison with Graph Methods (RQ3)}
We also compare the proposed Transition-Aware Embedding Distillation (TED) module with graph-based regularization methods in Table~\ref{tab:ablation}. The results show that most of the methods can improve the performance of MQSA. Specifically, GES performs better than GraReg on Amazon datasets but worse on the Yelp dataset. Moreover, our proposed TED method outperforms GES and GraReg in most cases, indicating the effectiveness of learning global and accurate item transition patterns by knowledge distillation.

\subsection{TED for Various Base Models (RQ4)}
We also compare the performance of various base models with and without our proposed Transition-Aware Embedding Distillation (TED) module in Table~\ref{tab:ITD}. The results demonstrate that TED can act as a domain adapter, which enhances the performance of the non-sequential method LightGCN on sequential recommendation tasks. Furthermore, the incorporation of TED yields remarkable improvement for the state-of-the-art sequential recommendation method FMLP-Rec. Notably, TED shows limited effects on the Yelp dataset due to the weak sequentiality of user interactions. In other words, transitional signals are less important in this dataset.

\subsection{Performance Comparison by Groups (RQ5)}
Figure~\ref{fig:group_exper} presents the performance of different methods on test samples grouped by transition frequencies observed in the training data from the validation item (the second last item) to the test item (the last item). We evaluate the SASRec model with the Transition-Aware Embedding Distillation (SASRec-TED), the Multi-Query Self-Attention model (MQSA), and the full MQSA-TED model. Compared with the results in Figure~\ref{fig:group_intro}, the improvement of MQSA over SASRec mainly results from the improvement on test samples lacking transition instances. However, the integration of long-query self-attention may hurt the performance on test samples with frequent transitions. By incorporating the TED module as a calibrator, MQSA-TED performs better than MQSA mainly on test samples with high transition frequencies. As MQSA and TED focus on collaborative and transitional signals, respectively, their combination will result in a reasonable balance between the two signal types.

\section{Related Work}

\textbf{Sequential Recommendation.} Sequential recommendation methods aim to capture dynamic user preferences \cite{he2017translation,chen2018sequential,wang2020make}. Early efforts adopt Markov Chains (MCs) to learn item transition patterns, such as FPMC \cite{rendle2010factorizing}, which combines the Matrix Factorization (MF) with the first-order Markov chain. Fossil \cite{he2016fusing} fuses the similarity-based model with high-order Markov chains. Recent efforts incorporate deep models, such as GRU4Rec \cite{hidasi2015session} and NARM \cite{li2017neural}, which employ Gated Recurrent Units (GRU). Caser \cite{tang2018personalized} uses horizontal and vertical convolutional filters. SASRec \cite{kang2018self} and BERT4Rec \cite{sun2019bert4rec} use unidirectional and bidirectional self-attention modules in Transformer \cite{vaswani2017attention}, respectively. FMLP-Rec \cite{zhou2022filter} is an all-MLP model with learnable filters in the frequency domain. However, previous efforts typically follow an auto-regressive framework, which neglects the valuable information in global item transition patterns. In this paper, we propose a Transition-Aware Embedding Distillation module to memorize and leverage the transitional signals.

\textbf{Self-Attention in Recommendation.} The Transformer architecture has achieved remarkable success in modeling long-term dependencies in Natural Language Processing (NLP) \cite{vaswani2017attention,devlin2018bert}. Consequently, recent efforts employ such architecture for sequential recommendation tasks \cite{ren2020sequential,qiu2022contrastive}. For example, SASRec \cite{kang2018self} and BERT4Rec \cite{sun2019bert4rec} use unidirectional and bidirectional self-attention modules, respectively. In addition, some efforts aim to enhance self-attention-based models by incorporating side information. For instance, TiSASRec \cite{li2020time} incorporates time interval embeddings into SASRec. S$^3$-Rec \cite{zhou2020s3} introduces self-supervision tasks to learn correlations among attributes, items, sub-sequences, and sequences based on mutual information maximization. SASRec-GES \cite{zhu2021graph} employs graph convolutions on sequential and semantic item graphs to generate smoothed item embeddings. Efforts have also been made to improve the efficiency or effectiveness of SASRec \cite{li2021lightweight}. CL4SRec \cite{xie2022contrastive} uses contrastive learning to derive self-supervision signals from user interaction sequences. Despite these advances, previous studies paid less attention to the limitations of the conventional self-attention architecture in capturing collaborative signals. In this paper, we propose a Multi-Query Self-Attention method that combines long and short-query self-attentions to enhance its effectiveness in modeling user collaborations.

\textbf{Knowledge Distillation in Recommendation}. Knowledge distillation is a widely-used model compression technique in various fields \cite{hinton2015distilling}, where a student model is trained with both a ground-truth label distribution and a smoothed pseudo-label distribution generated by a teacher model. Recent efforts apply this method to recommender systems, such as Ranking Distillation \cite{tang2018ranking}, which trains a student model to rank items based on both training data and teacher model predictions. Collaborative Distillation \cite{lee2019collaborative} uses probabilistic rank-aware sampling with teacher-guided and student-guided training strategies. Other existing methods aim to distill knowledge from side information into recommendation models to enhance their performance and interpretability. For instance, SCML \cite{zhu2020social} combines the item-based CF model with the social CF model through embedding-level and output-level mutual learning. DESIGN \cite{tao2022revisiting} integrates information from the user-item interaction graph and the user-user social graph and makes them learn from each other. Zhang et al. \cite{zhang2020distilling} propose a joint learning framework to distill structured knowledge from a path-based model into a neural model. 
However, knowledge distillation has received less attention in the context of sequential recommendation. In this paper, we distill the knowledge of item transitions into sequential recommendation models to enhance their performances.

\section{Conclusion}
In this paper, we addressed the limitations of existing sequential recommendation methods in capturing collaborative and transitional signals in user interaction sequences. To overcome these limitations, we proposed a new method called Multi-Query Self-Attention with Transition-Aware Embedding Distillation (MQSA-TED). To capture collaborative signals, we introduced an $L$-query self-attention module using flexible window sizes for attention queries and combined long and short-query self-attentions. In addition, we developed a transition-aware embedding distillation module that distills global item transition patterns into item embeddings, enabling the model to memorize and leverage transitional signals. Experimental results on four real-world datasets demonstrated the effectiveness of both modules in improving sequential recommendation performance.

\begin{acks}
  This research was partly supported by a CIHR-NSERC-SSHRC Healthy Cities Research Training Platform grant of Canada.
\end{acks}

\bibliographystyle{ACM-Reference-Format}
\balance
\bibliography{main}










\end{document}